\documentclass[trackchanges,resetfootnote, twocolumn]{aastex701}

\usepackage{amsmath}
\usepackage{booktabs}

\begin{document}

\title{Long-term Orbital Period Variations of the Eclipsing Dwarf Nova HT Cas}

\author[0000-0003-1399-5804]{Aykut Ozdonmez}
\affiliation{Atatürk University, Faculty of Science, Department of Astronomy and Space Science, 25240, Yakutiye, Erzurum, T\"urkiye}
\email[show]{aykut.ozdonmez@atauni.edu.tr}

\author[0000-0003-1339-3045]{Huseyin Er}
\affiliation{Atatürk University, Faculty of Science, Department of Astronomy and Space Science, 25240, Yakutiye, Erzurum, T\"urkiye}
\email{huseyin.er@atauni.edu.tr} 

\author[0000-0001-8131-4455]{Ilham Nasiroglu}
\affiliation{Atatürk University, Faculty of Science, Department of Astronomy and Space Science, 25240, Yakutiye, Erzurum, T\"urkiye}
\email{inasir@atauni.edu.tr}

\author[0000-0002-2847-8124]{M. Emir Kenger}
\affiliation{Atatürk University, Graduate School of Natural and Applied Sciences, Department of Astronomy and Astrophysics, 25240, Yakutiye, Erzurum, T\"urkiye}
\affiliation{Türkiye National Observatories, TUG, 07070 Antalya, Türkiye}
\email{emirkngr@gmail.com} 

\author[0009-0003-2863-5577]{Murat Tekkesinoglu}
\affiliation{Atatürk University, Graduate School of Natural and Applied Sciences, Department of Astronomy and Astrophysics, 25240, Yakutiye, Erzurum, T\"urkiye}
\email{murattekkesinoglu@gmail.com} 

\author[0000-0001-5778-5679]{Ergun Ege}
\affiliation{Istanbul University, Faculty of Science, Department of Astronomy and Space Sciences, 34116, Beyazit, Istanbul, T\"urkiye}
\email{ergunege@istanbul.edu.tr} 

\author[0000-0003-4120-0562]{Burak Batuhan Gürbulak}
\affiliation{Atatürk University, Graduate School of Natural and Applied Sciences, Department of Astronomy and Astrophysics, 25240, Yakutiye, Erzurum, T\"urkiye}
\email{xxxx@xxx.com} 

\author[0000-0003-1423-5516]{Ali Takey}
\affiliation{National Research Institute of Astronomy and Geophysics (NRIAG), 11421 Helwan, Cairo, Egypt}
\email{ali.takey@nriag.sci.eg} 

\author[0000-0003-3910-2285]{Nazlı Karaman}
\affiliation{Adiyaman University, Gölbaşı Vocational School, Department of Electricity and Energy, 02500, Gölbaşı, Adiyaman, T\"urkiye}
\affiliation{Adiyaman University, Astrophysics Application and Research Center, 02040, Adiyaman, T\"urkiye}
\email{nkaraman@adiyaman.edu.tr} 

\author[0000-0002-6878-760X]{M. Abdelkareem}
\affiliation{National Research Institute of Astronomy and Geophysics (NRIAG), 11421 Helwan, Cairo, Egypt}
\email{mohamed.abdelkareem@nriag.sci.eg}


\begin{abstract}
We present a comprehensive analysis of the long-term orbital period variations in the short-period eclipsing dwarf nova HT Cas. 
By combining our new high-precision mid-eclipse times obtained between 2015 and 2026 with archival data, we constructed an updated $O-C$ diagram spanning a $\sim$48-years.  
Statistical analysis confirms outbursts do not cause systematic phase shifts, validating the use of all activity states. Through MCMC modeling, we show that the $O-C$ variations require a two-companion configuration.
A free-eccentricity LTT model captures the variations but yields unconstrained posteriors and a highly eccentric outer orbit ($e_3 \sim 0.94$) that instantly collapses in N-body dynamical simulations. Imposing a circular constraint ($e=0$) resolves these mathematical degeneracies, yielding well-constrained posterior distributions. 
This dynamically stable model identifies two hypothetical circumbinary companions with minimum masses of $\sim 9.8 M_{Jup}$ and $\sim 5.0 M_{Jup}$, and periods of $\sim 32.6$ and $\sim 15.1$ years. Besides, this configuration inherently produces a negative quadratic term ($Q = -1.23 \times 10^{-14}$ days), aligning with secular period decrease predicted by standard CV evolution theory below the period gap.
Refined energy-budget tests reveal that classical Applegate mechanisms require significantly more energy than the secondary star provides, indicating they cannot independently drive the modulations. While advanced magnetic frameworks may offer theoretical alternatives, our findings demonstrate that a dynamically stable two-companion architecture provides a highly robust and physically viable explanation, consistent with second-generation planet formation within a post-common-envelope disk.
\end{abstract}



\section{Introduction}

Cataclysmic variables (CVs) are close interacting binaries whose long-term evolution is driven by angular momentum loss (AML) via gravitational radiation for short periods ($P_{orb} \leq 2\,\mathrm{h}$) and magnetic braking for longer periods \citep{1962ApJ...136..312K, 1981A&A...100L...7V, 2011ApJS..194...28K}.
In eclipsing CVs, high-time-resolution photometry enables precise mid-eclipse timing measurements. While secular period decreases are generally expected from standard AML, cyclic eclipse-timing variations (ETVs) often appear in $O-C$ diagrams. These modulations are typically attributed to either magnetic activity cycles in the secondary star (the Applegate mechanism; \citealt{1992ApJ...385..621A}) or the light-travel-time (LTT) effect induced by circumbinary companions \citep{2011A&A...526A..53B,2015MNRAS.448.1118G,2020ApJ...901..113F,2022NewA...9301751S}. 
However, ETV-based planetary hypotheses require independent validation. Many proposed multi-body configurations suffer from rapid dynamical instability on short timescales or are physically implausible \citep{2010MNRAS.404..837H, 2011MNRAS.416L..11H, 2024PASA...41...47E, 2025ARep...69..758K, 2025NewA..11902414E}.
Furthermore, energy-budget calculations often reveal that the Applegate mechanism demands more energy than the secondary star can physically provide \citep{2016A&A...587A..34V, 2020MNRAS.491.1820L}. Thus, comprehensive dynamical and energetic analyses are crucial to confirm the physical origin of these cyclic variations.

HT Cas is a well-studied, short-period eclipsing SU UMa-type dwarf nova \citep{1943AN....274...36H, 1979BAAS...11..664P, 1981ApJS...45..517P, 1986ApJ...305..740Z, 1987AN....308...75W}. Its fundamental parameters are well constrained, with a mass ratio of $q = 0.15 \pm 0.03$ ($M_{WD}=0.61 \pm 0.04\,M_\odot$ and $M_2=0.09 \pm 0.02\,M_\odot$) and an orbital period of $P_{orb} \sim 1.77$ h \citep{1991ApJ...378..271H}. It typically remains in a faint quiescent state ($V \approx 16.4$), displaying deep, stable 5-minute eclipses ideal for long-term timing analysis \citep{1981ApJ...245.1035Y}. However, the system exhibits notable photometric variability, including transitions between high and low quiescent states (15.9 and 17.7 mag; \citealt{1996AJ....112.2248R, 1999MNRAS.310..398I}) and occasional outbursts.  
While early studies found no significant orbital period variations using continuously updated mid-eclipse times \citep{1981ApJS...45..517P, 1986ApJ...305..740Z,1991ApJ...378..271H,1997ApJ...475..812M,1999MNRAS.310..398I,2000A&A...359..998B, 2005MNRAS.364.1158F}, \citet{2008A&A...480..481B} later identified a $\sim36$-year cyclic modulation. Recently, \citet{2023ApJ...953...63H} detected a secular period decrease combined with a 30.28-year cyclic oscillation. They proposed that standard AML mechanisms fail to explain the secular period decrease, suggesting empirical consequential AML (eCAML) instead, and attributed the cyclic signal to a highly eccentric ($e=0.82$) giant planet. 
However, the physical origins of such complex period modulations remain an open question in the literature.
Recent studies \citep{2024ApJ...966..155S, 2024ApJ...972...33S} confirmed the secular decrease but questioned the third-body hypothesis, arguing that timing data from the last decade do not support a strict periodic model and that magnetic activity is a statistically more likely driver.

In the paper, Section \ref{sec:obs} details our new photometric observations and the $\sim$48-year timing dataset. In Section \ref{sec:3}, we analyze the orbital period variations via Markov Chain Monte Carlo (MCMC) modeling of LTT scenarios and assess the outbursts impacts on eclipse timings. Section \ref{sec:4} evaluates the energetic viability of magnetic activity cycles. Section \ref{sec:5} tests the N-body dynamical stability of the proposed circumbinary configurations. Finally, conclusions are presented in Section \ref{sec:discussion}.

\section{Observations and Timing Data}
\label{sec:obs}
HT Cas was monitored between August 2015 and February 2026 using three telescopes: the 0.6 m telescope (ADYU60, Adıyaman, Türkiye) at the Adıyaman University Observatory, equipped with an Andor iKon-M 934 CCD camera ($13 \times 13\, \mu$m pixel size); the 1.0 m telescope (TUG100, Antalya, Türkiye) at the Türkiye National Observatories, equipped with a 4k $\times$ 4k SI1100 CCD camera ($15 \times 15\, \mu$m pixel size); and the 1.88 m telescope (088, Cairo, Egypt) at the Kottamia Astronomical Observatory (KAO), equipped with a 2k $\times$ 2k  E2V 42-40 CCD camera ($13.5 \times 13.5\, \mu$m pixel size) that is part of the Kottamia Faint Imaging Spectro-Polarimeter \citep[KFISP;][]{2022ExA....53...45A}. To maximize signal-to-noise ratio (S/N) and time resolution for precise determination of mid-eclipse times, observations were performed in the clear band with exposure times of 3 s for KAO, 2–5 s for TUG100, and 10–20 s for ADYU60.

To maximize the overall sampling efficiency by keeping readout times as short as possible, the KAO operated with $2 \times 2$ binning via dual amplifiers (yielding a 3 s readout), and the ADYU60 used its native fast readout ($\sim$1 s). For the TUG100, the readout time was successfully reduced to 2–3 seconds by utilizing a $300 \times 300$ pixel sub-array combined with the $2 \times 2$ binning mode. Standard data reduction procedures, including bias subtraction, dark current correction, and flat-fielding, were applied to all raw images. Differential aperture photometry was performed following the same methodology described by \citet{2021MNRAS.507..809E, 2023MNRAS.526.4725O, 2024PASA...41...47E, 2025ApJ...994...67E}. The photometric data were extracted using a Python script, utilizing a nearby non-variable comparison star to generate the eclipse light curves. In total, 83 new eclipse light curves were obtained from this ground-based monitoring campaign. In addition to our ground-based observations, mid-eclipse times were derived from the photometric data available in the American Association of Variable Star Observers (AAVSO) archives\footnote{https://www.aavso.org}.

\begin{figure}
\includegraphics[width=\columnwidth]{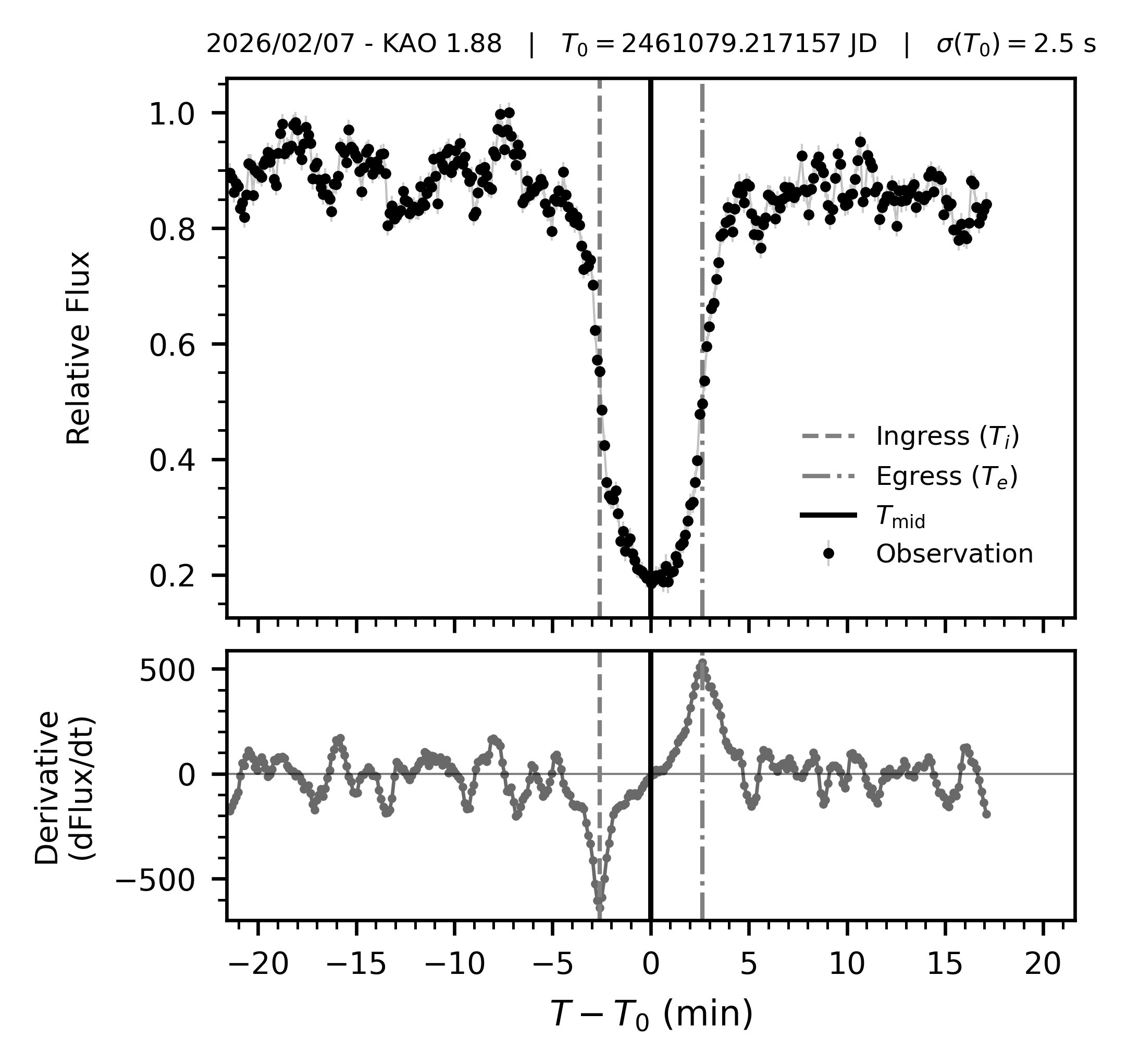}
\caption{
An example of determining the mid-eclipse time for HT Cas using the KAO 1.88 m telescope. The top panel shows the observed relative flux (black points). The bottom panel displays the corresponding derivative curve (solid gray line). The dashed and dash-dotted vertical lines refer to the mid-ingress ($T_i$) and mid-egress ($T_e$) times, respectively. The central solid vertical line denotes the resulting mid-eclipse time ($T_{mid}$).
}
\label{fig:LC}
\end{figure}

Although the eclipse profiles of HT Cas vary between high and low brightness states due to the accretion disk and bright spot activity, the ingress and egress times of the white dwarf remain stable \citep{2005MNRAS.364.1158F, 2023ApJ...953...63H}. This stability makes them a reliable reference for timing analysis. We used the derivative method described by \citet{1985MNRAS.214..475W} to measure the mid-eclipse times ($T_{mid}$). In this method, the times of mid-ingress ($T_i$) and mid-egress ($T_e$) are identified as the minimum and maximum points of the light curve’s derivative, respectively. The example eclipse light curve obtained from our observations is shown in Figure \ref{fig:LC}. The mid-eclipse times were calculated using the equation $T_{mid}=(T_i+T_e)/2$. Finally, all mid-eclipse times were converted to Barycentric Julian Date (BJD) \citep{2010PASP..122..935E}. 

\begin{figure}
\includegraphics[width=1\columnwidth]{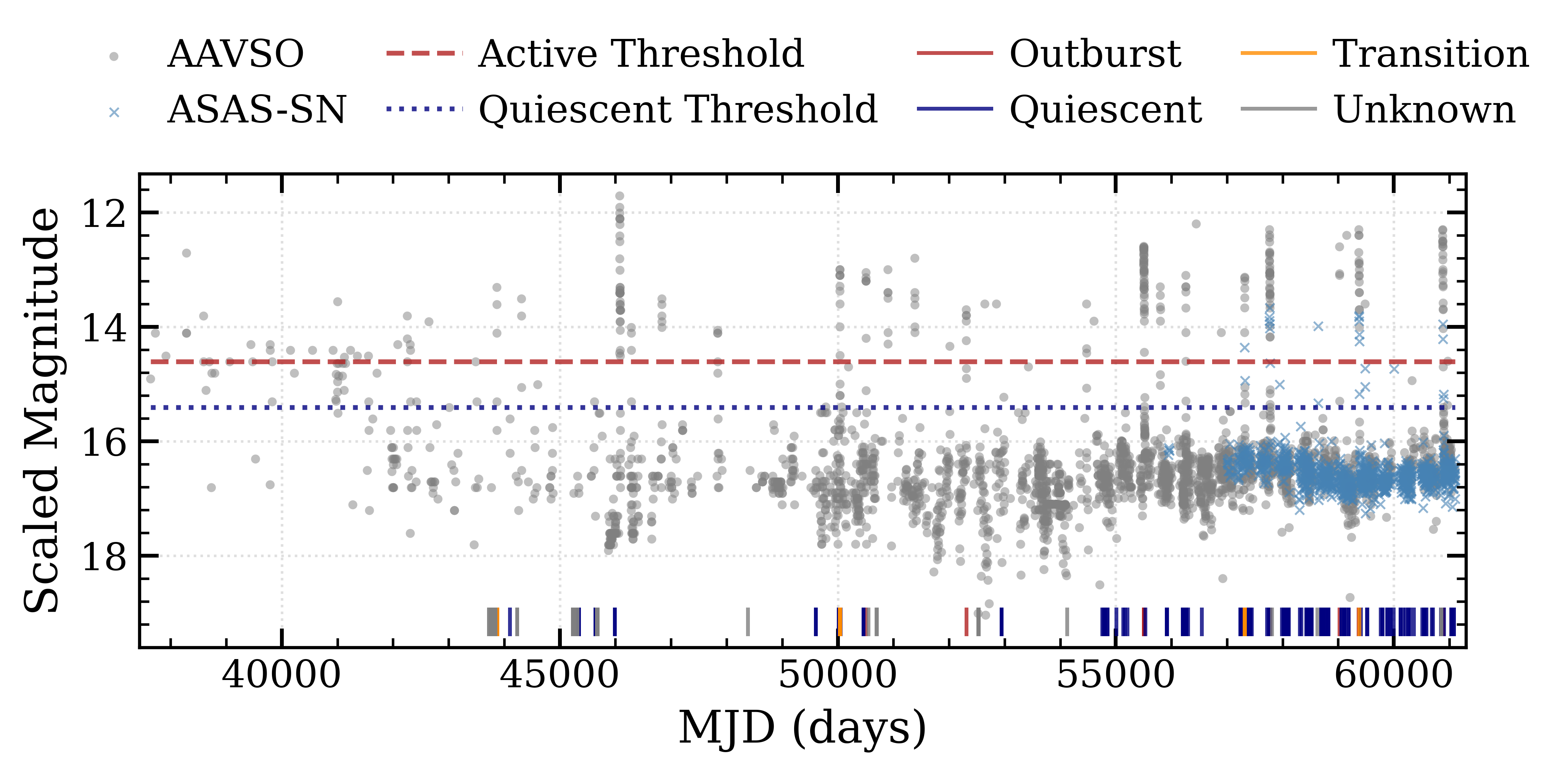}
\caption{The long-term scaled photometric light curve of HT Cas compiled from ASAS-SN and AAVSO databases. The horizontal dashed and dotted lines represent the dynamically calculated global outburst and quiescent thresholds, respectively. To demonstrate the distribution of the timing data across different activity states, the epochs of the measured mid-eclipse times are marked as short vertical colored lines (rug plot) at the bottom of the panel: Red vertical lines indicate the eclipse times during outbursts, blue during quiescence, orange during transitions, and gray for unknown states.} 
\label{fig:long_lc}
\end{figure}

To systematically identify the activity states of HT Cas and to cross-match these periods with timings from the literature and our own data, we analyzed the long-term light curves from the All-Sky Automated Survey for Supernovae (ASAS-SN) Variable Stars \citep{2014ApJ...788...48S,2020MNRAS.491...13J} and the AAVSO databases. A global auto-threshold calculation was applied to the combined long-term light curve data. The global baseline median was calculated as 16.61 mag (scaled), with a noise level of 0.402 mag ($1\sigma$). Based on these values, the global quiescent threshold was defined as 15.41 mag ($3\sigma$), and the global outburst threshold was defined as 14.60 mag ($5\sigma$). Brightness levels falling between these two limits were classified as the transition state. 
To assign an activity state to each specific $O-C$ data point, we employed a cross-matching procedure using a temporal window of $\pm 5$ days centered on each mid-eclipse time. The representative brightness for a given epoch was derived from the median magnitude of all photometric observations falling within this window. This local median was then evaluated against our established global thresholds. If the designated temporal window contained no long-term photometric data, the activity status of the corresponding eclipse was classified as "Unknown". 
Using this classification scheme, we determined the activity status of the system corresponding to the available mid-eclipse times in the $O-C$ data. Out of the 376 total $O-C$ data points, 247 were obtained during the quiescent state, 52 during the outburst state, 12 during the transition state, and 65 were classified as unknown (Table \ref{tab:midtimes_htcas}). The long-term light curves along with the determined thresholds and the corresponding $O-C$ data distribution are illustrated in Figure \ref{fig:long_lc}.

\section{Evaluation of Orbital Period Variations}
\label{sec:3}
\subsection{Eclipsing Times}
To comprehensively analysis of the orbital period variations of HT Cas, we constructed an updated $O-C$ diagram by combining our new mid-eclipse times with literature data. Mid-eclipse times for this system have been reported by several authors \citep{1981ApJS...45..517P,1986ApJ...305..740Z,1991ApJ...378..271H,1997ApJ...475..812M,1999MNRAS.310..398I,2000A&A...359..998B,2005MNRAS.364.1158F}. \citet{2008A&A...480..481B} constructed an $O-C$ diagram using eleven mid-eclipse times derived from annual averages of the literature data (22 mid-eclipse times excluding \citealt{2000A&A...359..998B}). \citet{2023ApJ...953...63H} significantly expanded the data set by measuring 31 new mid-eclipse times and deriving 92 times from the AAVSO database. However, their analysis excluded 23 mid-eclipse times from \citet{1999MNRAS.310..398I}, which cover both quiescence and outburst states. Recently, \citet{2024ApJ...966..155S} investigated the secular period change using literature data compiled by \citet{2008A&A...480..481B}, including mid-eclipse times derived from phase-folded AAVSO and Zwicky Transient Facility (ZTF) light curves, but did not incorporate the recent mid-eclipse times reported by \citet{2023ApJ...953...63H}. 

We compiled a complete data set that bridges these gaps, incorporating all available historical data, the most recent literature values, and our new observations to ensure the most comprehensive analysis to date. We also included mid-eclipse times from the VarAstro\footnote{\url{https://var.astro.cz/}} project of the Variable Star and Exoplanet Section of the Czech Astronomical Society \citep{2006OEJV...23...13P}, neither of which has been used in previous orbital period studies of this system. 
Before investigating the orbital period variation of HT Cas, several pre-processing steps were applied to the compiled mid-eclipse times to ensure a high-quality and homogeneous dataset. First, we excluded the ZTF timing data published by \citet{2024ApJ...966..155S}, as the insufficient temporal resolution of these phase-folded light curves does not meet the precision requirements for determining reliable mid-eclipse times. We also restricted our analysis exclusively to primary eclipses to maintain homogeneity; for instance, three secondary minima identified among the eight timings in the VarAstro database were removed. 
To ensure high-precision timing, we excluded archival AAVSO light curves with large uncertainties, incomplete eclipse profiles, or low temporal cadence, yielding 51 mid-eclipse times for our orbital analysis. Additionally, six archival timings \citep{1999MNRAS.310..398I, 2000A&A...359..998B, 2023ApJ...953...63H} exhibiting significant and inconsistent scatter relative to the general $(O-C)$ trend were excluded as outliers during LTT modelling.
Based on the distribution of existing valid errors, we calculated a median value of $0.000100$ d and a standard deviation ($\sigma$) of $0.000065$ d. To reduce bias while retaining these points, we assigned a $+3\sigma$ limit of $0.000295$ d as the standard error for 104 data points that lacked valid error estimates in the literature. 

While \citet{2023ApJ...953...63H} used archival data from the AAVSO to extend their $O-C$ analysis until 2021 June, our study extends this timeline by nearly five years, incorporating new high-precision observations up to 2026 February. By minimizing observational gaps through a combination of our ground-based campaigns and archival databases, we provide an updated, sampled $\sim$48-year observational baseline for HT Cas. All mid-eclipse times are listed in Table \ref{tab:midtimes_htcas}. 

\begin{table}[ht]
\centering
\caption{Mid-eclipse times, uncertainties, references and derived activity status for HT Cas. }
\label{tab:midtimes_htcas}
\begin{tabular}{lccc}
\hline
BJD & Error &  References & Activity Status\\
\hline
\multicolumn{4}{c}{From AAVSO} \\ 
\hline
2452312.32964070 & 0.00020209 & AAVSO & Outburst\\
2455503.46152282 & 0.00018293 & AAVSO & Outburst\\
2455503.53534059 & 0.00018730 & AAVSO & Outburst\\
... & ... & ...\\
\hline
\multicolumn{4}{c}{From our data} \\
\hline
2457238.51584465 & 0.00004437 &  ADYU60 & Quiescent\\
2457719.43210012 & 0.00004242 &  TUG100 & Quiescent\\
2461033.18814982 & 0.00002676 &  KAO & Quiescent\\
... & ... & ...\\
\hline
\end{tabular}
\tablecomments{This table is available in its entirety in machine-readable form in the online article.}
\end{table}

\subsection{Secular Variations: The Linear and Quadratic Trend}
To investigate potential long-term trends in the orbital period of HT Cas, we tested both linear and quadratic ephemeris fits to the updated data.
We derived the linear ephemeris parameters using a weighted least-squares fit to all available mid-eclipse times, as follows:

\begin{equation}
\begin{split}
\label{eq:lineer_eph}
T_{eph}(E) &= T_0 + P_{bin} \times E \\
   &=  \text{BJD}\: 2443727.938351 (71) \\
    & \quad + 0.0736471945 (3) \times E \\
\end{split}
\end{equation}

Here, $T_{eph}(E)$ is the BJD of the mid-eclipse at epoch $E$, while $T_0$ and $P_{bin}$ are the initial ephemeris and the orbital period, respectively.

The weighted least-squares fit yielded a statistically significant quadratic ephemeris ($p \approx 2.68 \times 10^{-4}$), defined as follows:
\begin{equation}
\begin{split}
\label{eq:quad_eph}
T_{eph}(E) &= T_0 + P_{bin} \times E + Q E^2 \\
    &= \text{BJD}\: 2443727.938635(104) \\
   & \quad + 0.0736471893(14) \times E \\
   & \quad + 1.846(50) \times 10^{-14} \times E^2
\end{split}
\end{equation}

Here, $Q$ is the quadratic term given by $P\dot{P}/2$, where $\dot{P}$ is the derivative of the orbital period with respect to time ($dP/dt$).
The inclusion of a quadratic term provided a statistically better fit than the linear model, reducing the Bayesian Information Criterion (BIC) by $\Delta \text{BIC} = -7.5$. This simple quadratic fit implies a secular orbital expansion with a orbital period change rate of $\dot{P} \simeq 5.01 \times 10^{-13}$. However, this contradicts the standard evolutionary theory of short-period CVs below the period gap, where AML via gravitational radiation should force the orbit to shrink ($Q < 0$).
Notably, fitting a simple parabola to a dataset containing long-term, incomplete cyclical variations often yields an artificial period derivative. Therefore, we do not adopt this positive quadratic ephemeris as a standalone physical solution, but rather utilize it as a mathematical baseline component for the comprehensive multi-periodic LTT modeling.

\subsection{LTT Solutions via MCMC Modeling}

To explain the cyclic variations, we considered the LTT effect caused by the gravitational perturbation of hypothetical circumbinary companion(s) \citep{1952ApJ...116..211I}. By combining the long-term parabolic trend with the multi-periodic LTT modulations, the model for fitting the $O-C$ variations is defined as:

\begin{equation}
\label{eq:OC}
O-C\: \text{[days]}= Q E^2 + \sum_{i} \tau_{i}(E)
\end{equation}

The LTT formulation ($\tau_{i}$) for each hypothetical companion, originally derived by \citet{1952ApJ...116..211I} and later modified by \citet{2012MNRAS.425..930G}, is defined as: 

\begin{equation}
\label{eq:tau}
\tau_{i}= K_i \left[\sin{\omega_i} (\cos{E_i (t)} - e_i) + \sqrt{1-e_i^2} \cos{\omega_i} \sin{E_i (t)}\right]
\end{equation}

In this expression, the subscript $i$ denotes the $i$-th companion orbiting the center of mass of the binary. $K_i$ represents the semi-amplitude of the LTT signal, while $e_i$ and $\omega_i$ correspond to the orbital eccentricity and the longitude of the pericenter, respectively. $E_i$ is the eccentric anomaly, and $t_{0,i}$ is the time of pericenter passage. To prevent weakly constrained solutions for $e_i$ and $\omega_i$, we implemented Poincaré orbital elements defined as $x_i \equiv e_i \cos\omega_i$ and $y_i \equiv e_i \sin\omega_i$, allowing for a more robust exploration of the parameter space \citep{2012MNRAS.425..930G, 2017AJ....153..137N, 2023MNRAS.526.4725O, 2025ApJ...994...67E}.

To determine the posterior distributions of the model parameters and assess their uncertainties, we employed a MCMC approach, following the Bayesian framework established in our previous studies \citep{2021MNRAS.507..809E, 2025ApJ...994...67E, 2023MNRAS.526.4725O}. We adopted uniform priors for all parameters, with boundaries chosen to cover a wide, physically plausible parameter space. Specifically, the search was restricted to positive values for $K_i, P_i, t_{0,i}$, and $\sigma_f$. The Poincaré elements ($x_i, y_i$) were constrained within the range $[-0.99, +0.99]$ to maintain numerical stability. In addition, $P_{bin}$ and $T_0$ were tightly constrained, allowed to vary by only a few days around their initial ephemeris values.

To account for unmodeled noise sources and potentially underestimated observational uncertainties, a systematic uncertainty parameter ($\sigma_f$) was added to the likelihood function. This parameter was added in quadrature to the formal mid-eclipse timing uncertainties ($\sigma_i^2 \rightarrow \sigma_i^2 + \sigma_f^2$) and treated as a free parameter, ensuring that the reduced chi-square approximates unity and provides robust posterior distributions, as described in our earlier works \citep[e.g.,][]{2021MNRAS.507..809E, 2025ApJ...994...67E, 2023MNRAS.526.4725O}. To sample the posterior probability distributions, we utilized the affine-invariant ensemble sampler provided by the \textit{emcee} package \citep{2013PASP..125..306F}. The final MCMC simulations for each model configuration were run using 256 walkers, each evolving over 180,000 iterations to ensure convergence.

\subsection{Evaluation of Unconstrained and Circular LTT Solutions}

\begin{figure*}
\centering
\includegraphics[width=0.85\textwidth]{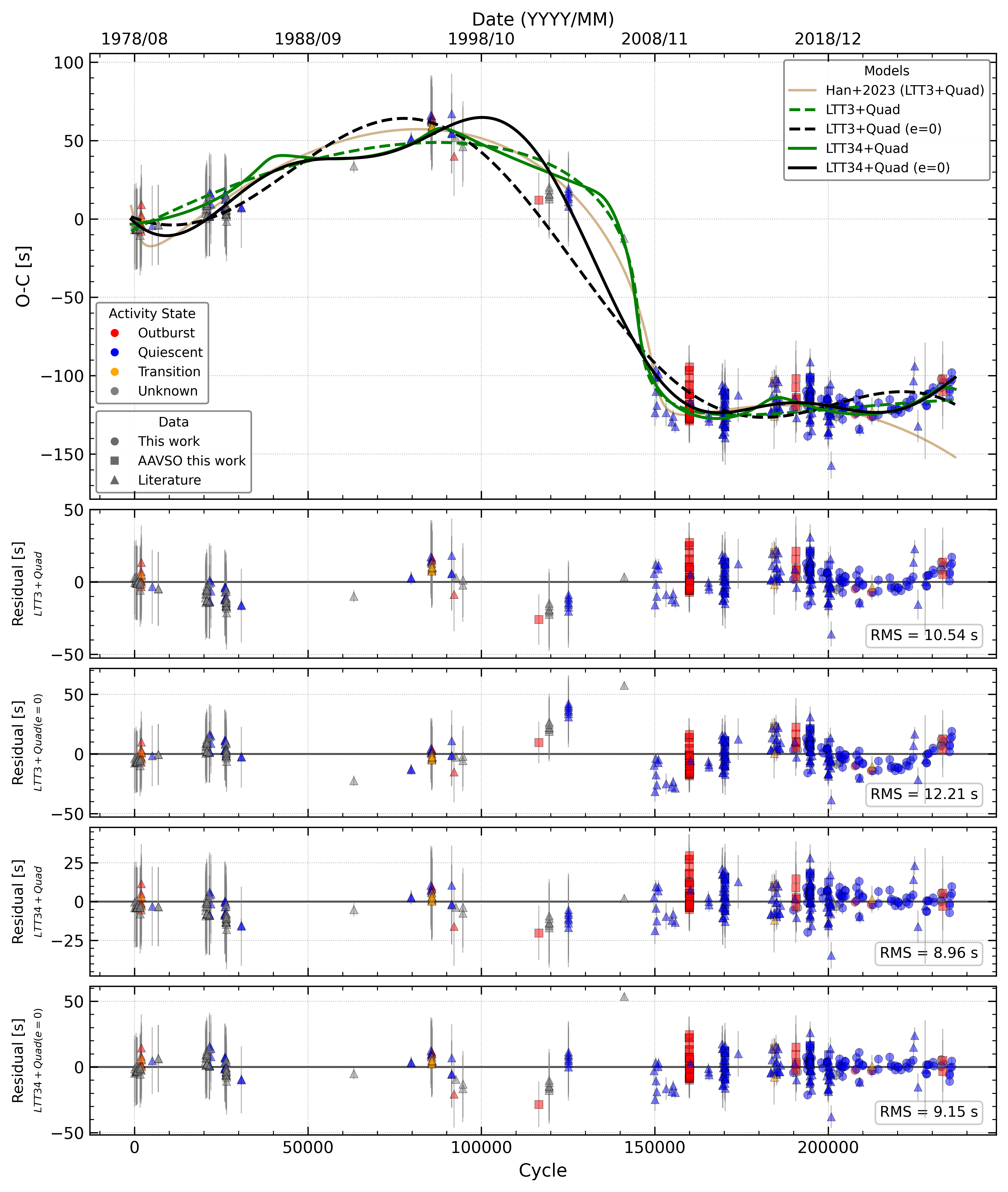}
\caption{The $O-C$ diagram of HT Cas and the corresponding residuals for the evaluated LTT models. Top panel: The primary $O-C$ data of HT Cas calculated with respect to the previous linear ephemeris for comparison; i.e. for $T_0=2443727.937862$ and $P_{bin}=0.0736472031$ days \citep{2023ApJ...953...63H}. The theoretical $O-C$ curves of the literature model \citep[][light brown solid line]{2023ApJ...953...63H} and our updated models including LTT3+Quad (green dashed line), LTT34+Quad (green solid line), LTT3+Quad ($e=0$, black dashed line) and LTT34+Quad ($e=0$, black solid line), are superimposed. Lower panels: The residuals of the eclipse timings after subtracting the LTT3+Quad, LTT3+Quad ($e=0$), LTT34+Quad, and LTT34+Quad ($e=0$) models, respectively. The RMS values for each fit are given in the corresponding panels. The marker symbols indicate the data sources: circles for our new observations, squares for AAVSO data from this work, and triangles for the literature data. The colors of the data points represent the activity state of the dwarf nova during observation: red (outburst), blue (quiescent), orange (transition), and gray (unknown).} 
\label{fig:o-c}
\end{figure*}

To find the most physically and statistically consistent representation of the $O-C$ data, we tested four different LTT models including both free-eccentricity and circular orbits for single and double companions. The derived parameters for the best-fit models are summarized in Table \ref{tab:htcas_models}, and the corresponding $O-C$ curves and residuals for all models are presented in Fig. \ref{fig:o-c}.

\begin{figure}
\includegraphics[width=\linewidth]{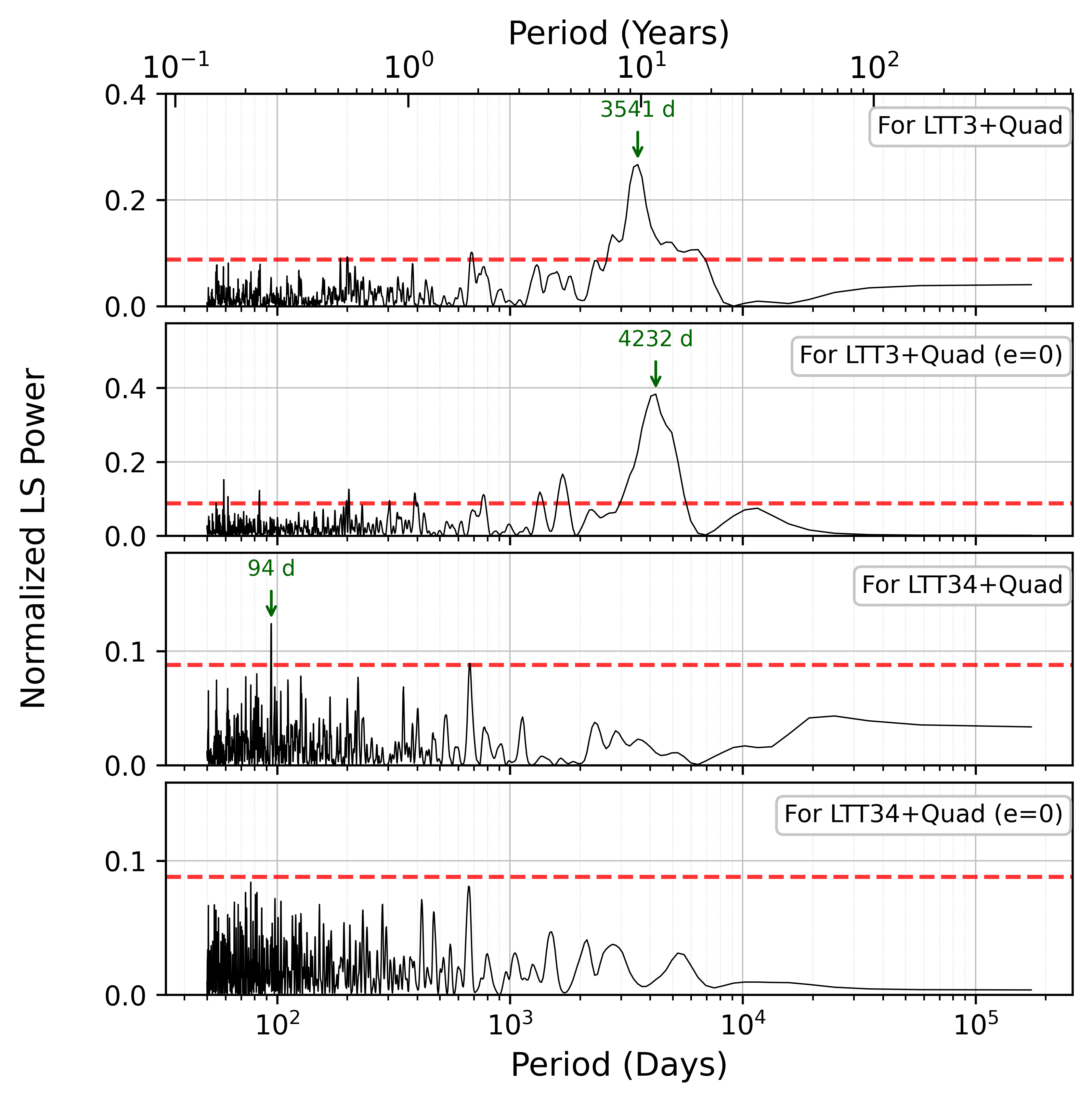}
\caption{The Lomb-Scargle periodograms of the $O-C$ residuals corresponding to the LTT3+Quad, LTT3+Quad ($e=0$),  LTT34+Quad, and LTT34+Quad ($e=0$) models (from top to bottom). The horizontal dashed red line in each panel represents the 0.01\% FAP significance level. Green arrows mark the centers of the significant periodic signals exceeding this threshold, with their corresponding periods labeled.} 
\label{fig:ls_for_residuals}
\end{figure}

To avoid any a priori bias and explore the physical parameter space, we initially allowed all orbital parameters, including eccentricities, to vary freely. First, we evaluated the \textit{LTT3+Quad} model, which assumes a single circumbinary companion with free eccentricity superimposed on the quadratic trend. The MCMC algorithm converged to a solution with a root-mean-square (RMS) scatter of 10.54 s. The derived orbital period of the third body is $P_3 \sim 153.2$ years with a semi-amplitude of $K_3 \sim 379.2$ s. However, the resulting orbit is so highly eccentric that it is almost parabolic, and dynamically improbable ($e_3 \sim 0.96$). In addition, a visual inspection of the residuals (Fig. \ref{fig:o-c}) reveals a distinct, unmodeled cyclic structure. To quantify this, we performed a Lomb-Scargle (LS) periodogram analysis \citep{1976Ap&SS..39..447L, 1982ApJ...263..835S} using the \texttt{astropy} package \citep{2022ApJ...935..167A} on the residuals of the \textit{LTT3+Quad} model, as in our recent studies \citep[e.g.,][]{2025AdSpR..76.1204E, 2026AdSpR..77.1365E}. As shown in the top panel of Fig. \ref{fig:ls_for_residuals}, the periodogram exhibits a broad and significant peak spanning a range of $7-20$ years, centered at $\sim 10$ years. This prominent peak far exceeds the $0.01\%$ False Alarm Probability (FAP) threshold, indicating the presence of an additional, unmodeled periodic signal.

Motivated by the LS periodogram results, we introduced a fourth body to construct the \textit{LTT34+Quad} model with free Poincaré orbital elements (i.e. eccentricities). This two-companion model resolved the residual periodicity, reducing the RMS scatter to 8.96 s and yielding a significantly improved $\Delta BIC$ of $-48.2$ compared to the reference single-companion model. While the dominant long-term periodicity was removed, the LS periodogram of the residuals (Fig. \ref{fig:ls_for_residuals}) revealed a sharp, high-frequency peak at approximately 94 days. 
Given the temporal distribution of our data (mean observation gap of 46.3 days and a maximum gap of $\sim 2390$ days), the detected 94-day peak is nearly twice the mean sampling interval, and it coincides directly with this theoretical limit, indicating that it is a spectral alias (i.e., a pseudo-Nyquist period) rather than a genuine physical modulation.
Thus, we do not attribute this signal to a fifth circumbinary body.

\begin{figure*}
    \centering
    \includegraphics[width=\textwidth]{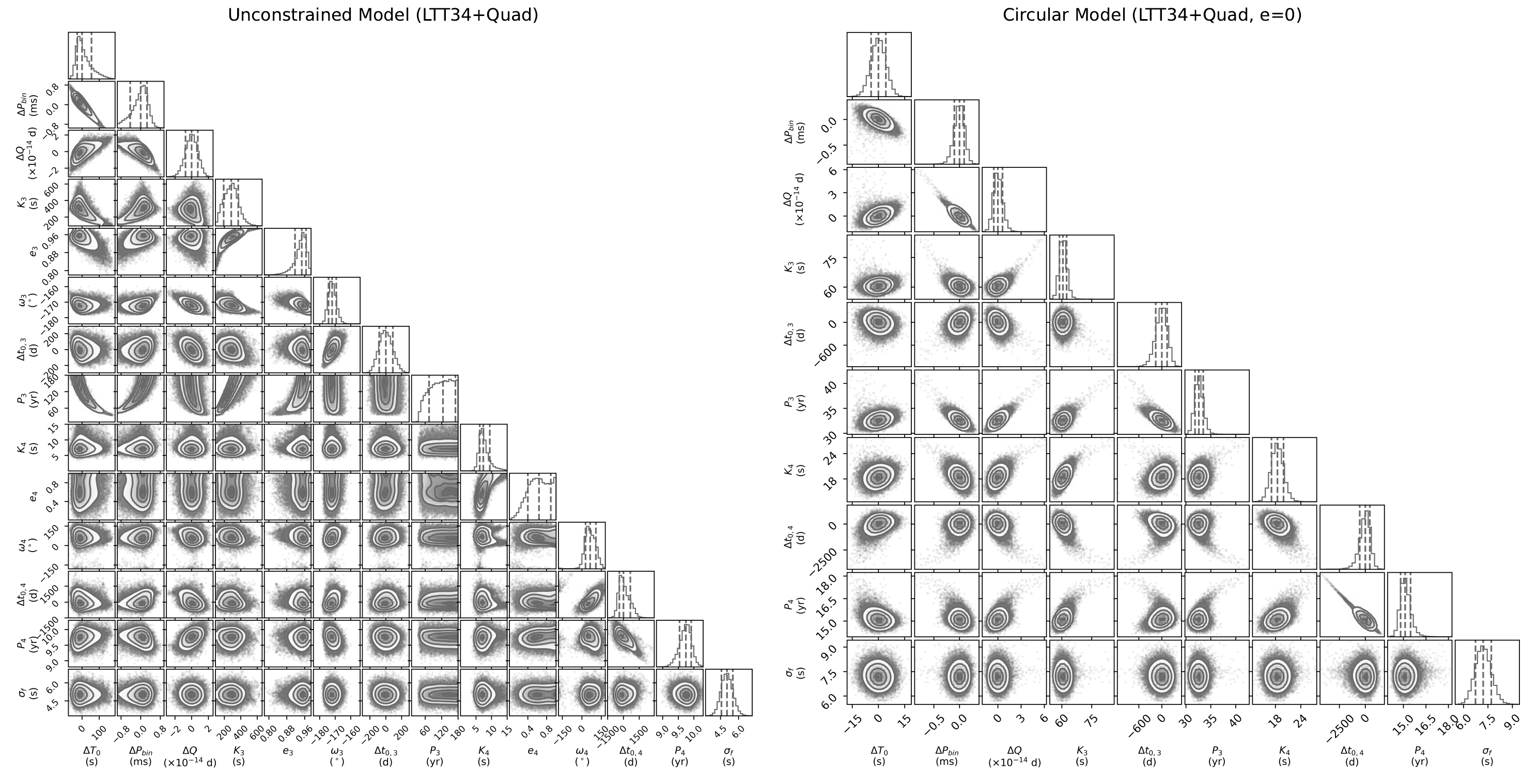}
    \caption{The 1D and 2D posterior probability distributions derived from the MCMC analysis of the two-companion models for HT Cas. \textit{Left:} The unconstrained model (LTT34+Quad). \textit{Right:} The circular model (LTT34+Quad, $e=0$). For optimal visualization, the parameters $T_0$, $P_{bin}$, $Q$, $t_{0,3}$, and $t_{0,4}$ are presented as differential values ($\Delta$) relative to their respective best-fit values. The vertical dashed lines in the 1D histograms indicate the $16^{th}$, $50^{th}$, and $84^{th}$ percentiles.}
    \label{fig:corner_plots}
\end{figure*}

Although the free-eccentricity \textit{LTT34+Quad} model provides the lowest (best) $\Delta BIC$ value mathematically, this configuration raises physical concerns from a dynamical perspective. The model identifies two companions with highly eccentric orbits ($e_3 \sim 0.94$ and $e_4 \sim 0.51$) and yields a massive outer companion of $M_3 \sin i_3 \sim 53.5\, M_{Jup}$. In a compact post-common-envelope binary (PCEB) system, hosting multiple massive substellar companions with such high eccentricities would likely lead to strong interactions, raising severe concerns regarding the physical validity of the orbital configuration \citep[e.g.][]{2010MNRAS.407.2362P,2011MNRAS.416.2202P,2012MNRAS.427.2812H,2012AJ....144...34H, 2013MNRAS.435.2033H}. Additionally, the MCMC chains for the $P_3$ parameter in the unconstrained models did not properly converge. Instead of forming a well-defined Gaussian peak, the posterior distribution diverges towards an unconstrained upper limit, resulting in orbital periods that far exceed the $\sim48$-year observational baseline (Fig. \ref{fig:corner_plots}, left panel). This indicates that the extreme geometric parameters ($e$ and $K$) yielded by the algorithm do not reflect a true physical orbit, but are rather a mathematical consequence of overfitting the sharp $O-C$ variations. Such spurious solutions are a known mathematical response to sparse data coverage (observational gaps) and the timing noise inherent to active CVs, especially when evaluated over a limited baseline.

To investigate whether the observed variations could be explained by a more favorable system configuration, we restricted the models by forcing circular orbits ($e=0$). First, we tested a single-companion circular model (\textit{LTT3+Quad, $e=0$}); however, it yielded a poor fit with an RMS scatter of 12.21 s and a $\Delta BIC$ penalty of $+133.3$ relative to the unconstrained \textit{LTT3+Quad} reference model. 
It failed to describe the variations in the $O-C$ diagram, leaving significant, unmodeled periodicities in the residuals (Fig. \ref{fig:ls_for_residuals}). Thus, we applied the circular constraint to the two-companion system, leading to the \textit{LTT34+Quad ($e=0$)} model.
Despite the reduction in degrees of freedom, the strictly circular two-companion model converged, yielding an RMS of 9.15 s. Although its higher $\Delta BIC$ ($+42.4$) indicates a nominally weaker statistical fit than the unconstrained \textit{LTT34+Quad} model, this circular configuration provides a physical improvement and yields well-constrained posterior distributions (Fig. \ref{fig:corner_plots}, right panel). In this circular configuration, the MCMC identified two fundamental periodicities: an outer companion with $P_3 = 32.6 \pm 0.8$ years and an inner companion with $P_4 = 15.1 \pm 0.3$ years.

The circular constraint resolves the extreme mass overestimation observed in the free-eccentricity model. In Keplerian modeling, an unconstrained highly eccentric orbit ($e \sim 0.95$) produces a strongly skewed LTT signal. To account for sharp variations, the MCMC algorithm required a large amplitude of  $K_3 \sim 282.7$ s, resulting in a minimum mass to $53.5\, M_{Jup}$ for the outer companion. However, restricting the system to a sinusoidal form ($e=0$) resolves this unphysical solution by eliminating the parameter divergence associated with highly eccentric projections. As seen in Table \ref{tab:htcas_models}, the circular assumption restricts the amplitude to $K_3 \sim 60.5$ s, bringing the derived mass down to a more realistic planetary regime ($M_3 \sin i_3 \sim 9.8\, M_{Jup}$). This demonstrates that the circular assumption prevents the unrealistic extreme orbital parameters of unconstrained eccentric fits, yielding physically robust mass constraints.

Our comprehensive analysis demonstrates that a single-body LTT model is statistically and observationally insufficient to explain the complex and multi-periodic $O-C$ variations of HT Cas. 
By flattening the residual periodograms and avoiding the mathematical degeneracy inherent to unconstrained eccentric fits, the adoption of a two-companion model (\textit{LTT34+Quad, e=0}) provides the most consistent interpretation of the system's long-term timing behavior. Unlike previous models where derived periods far exceeded the observational span, our adopted circular model constrains the outermost orbital period to $P_3 \sim 32.6$ years.

It is also noteworthy that the inclusion of LTT signals in our MCMC models forces the quadratic term to revert to a negative value ($Q = -1.23^{+0.60}_{-0.54} \times 10^{-14}$ days). This corrects the anomalous positive trend observed in the initial simple parabolic fit. The physical implications of this correction, particularly regarding the secular evolution and AML of the CV system, are thoroughly evaluated in Section  \ref{sec:discussion}.

\begin{table*}
\centering
\caption{Orbital, mass, and magnetic activity parameters of HT Cas derived from the MCMC analysis of the $O-C$ diagram. The free-eccentricity (LTT34+Quad) and circular (LTT34+Quad, $e=0$) two-companion models are compared. Note: The $\Delta BIC$ value for each two-companion model is calculated relative to its respective single-companion reference model (i.e., LTT3+Quad and LTT3+Quad, $e=0$).}
\label{tab:htcas_models}
\begin{tabular}{llcc}
\toprule
\textbf{Parameter} & \textbf{Unit} & \textbf{LTT34+Quad} & \textbf{LTT34+Quad ($e=0$)} \\
\midrule
\multicolumn{4}{l}{\textit{Orbital Ephemeris}} \\
$T_0$ & BJD & $2443728.083594^{+0.000623}_{-0.000332}$ & $2443726.9812^{+0.000048}_{-0.000049}$ \\
$P_{bin}$ & days & $0.07364721 \pm 10^{-8}$ & $0.07364720 \pm 10^{-8}$ \\
$Q$ & days & $(-3.42^{+0.72}_{-0.74}) \times 10^{-14}$ & $(-1.23^{+0.60}_{-0.54}) \times 10^{-14}$ \\
\midrule
\multicolumn{4}{l}{\textit{Third Body ($LTT_3$)}} \\
$K_3$ & s & $282.7^{+85.0}_{-91.8}$ & $60.5^{+1.9}_{-1.7}$ \\
$P_3$ & years & $124.1^{+46.3}_{-50.0}$ & $32.6^{+0.9}_{-0.8}$ \\
$t_{0,3}$ & BJD & $2454399^{+88}_{-82}$ & $2447687^{+142}_{-159}$ \\
$e_3$ & & $0.945^{+0.020}_{-0.028}$ & - \\
$\omega_3$ & deg & $-171.9^{+2.6}_{-2.3}$ & - \\
$a_{3} \sin i_3$ & AU & $22.072^{+5.198}_{-6.427}$ & $9.062^{+0.206}_{-0.196}$ \\
$M_3 \sin i_3$ & $M_{Jup}$ & $53.5^{+24.2}_{-15.3}$ & $9.8^{+0.4}_{-0.4}$ \\
\midrule
\multicolumn{4}{l}{\textit{Fourth Body ($LTT_4$)}} \\
$K_4$ & s & $7.3^{+2.1}_{-1.1}$ & $18.5^{+1.5}_{-1.4}$ \\
$P_4$ & years & $9.7^{+0.2}_{-0.2}$ & $15.1^{+0.4}_{-0.3}$ \\
$t_{0,4}$ & BJD & $2439466^{+655}_{-390}$ & $2434129^{+440}_{-533}$ \\
$e_4$ & & $0.508^{+0.377}_{-0.109}$ & -\\
$\omega_4$ & deg & $64.6^{+43.6}_{-36.4}$ & - \\
$a_{4} \sin i_4$ & AU & $4.151^{+0.066}_{-0.069}$ & $5.451^{+0.087}_{-0.072}$ \\
$M_4 \sin i_4$ & $M_{Jup}$ & $2.9^{+1.6}_{-0.4}$ & $5.0^{+0.4}_{-0.4}$ \\
\midrule
\multicolumn{4}{l}{\textit{Statistics \& Goodness-of-Fit}} \\
$\sigma_f$  & s & $5.0^{+0.5}_{-0.5}$ & $7.2^{+0.5}_{-0.5}$ \\
RMS & s & $8.96$ & $9.15$ \\
$\Delta BIC$ & & $-48.19$ & $-139.02$ \\
\midrule
\multicolumn{4}{l}{\textit{Magnetic Activity Tests for $LTT_3$}} \\
Thin-shell $\Delta E/E_{sec}$ & & $0.038$ & $0.096$ \\
Two-zone $\Delta E/E_{sec}^-$ & & $0.68$ & $1.72$ \\
Spin-orbit coupling $\Delta E_{rot}/E_{2}$ & & $7.74$ & $23.98$ \\
\midrule
\multicolumn{4}{l}{\textit{Magnetic Activity Tests for $LTT_4$}} \\
Thin-shell $\Delta E/E_{sec}$ & & $0.053$ & $0.089$ \\
Two-zone $\Delta E/E_{sec}^-$ & & $0.94$ & $1.60$ \\
Spin-orbit coupling $\Delta E_{rot}/E_{2}$ & & $32.50$ & $34.02$ \\
\bottomrule
\end{tabular}
\end{table*}

\subsection{Dependency of Outburst Activity on Eclipse Timing}
In CVs, mid-eclipse times measured during outbursts or active states can sometimes exhibit systematic offsets. These deviations generally arise from the expansion of the accretion disk or the migration of the hot spot, which shifts the center of light during the eclipses \citep{1984AcA....34..161S, 1985MNRAS.213..129H, 2024ApJ...966..155S}. Therefore, it has become a standard methodology in orbital period studies to strictly exclude eclipse timings obtained during outbursts to avoid these potential timing offsets \citep[e.g.,][]{2008A&A...480..481B, 2023ApJ...953...63H}. Although this filtering approach is widely adopted, quantitative and systematic analyses evaluating the actual impact of including non-quiescent data on the $O-C$ diagrams and the resulting orbital parameters are rarely performed. To investigate this effect, we classified the available mid-eclipse times according to the activity states of HT Cas, using the long-term photometric data and the threshold criteria detailed in Section \ref{sec:obs}.

To quantitatively evaluate the immediate impact of outbursts on individual timings, we first performed a model-independent local reference analysis, focusing on the clearly defined quiescent and outburst data (excluding transition and unknown states). For each timing recorded during an active state, we established a local quiescent reference using the immediately surrounding quiescent data points (within a $\pm 2000$ cycle window). To ensure the robustness of our statistics against unphysical observational artifacts, we filtered extreme outliers (deviations $> 15$ minutes, which excluded only a single anomalous data point) and utilized median-based metrics. 
Our non-parametric test (Mann-Whitney U) reveals that the median timing deviation during outbursts is merely $\sim 3.6$ seconds compared to local quiescent neighbors. This difference yields a p‑value of $0.067$, which is greater than the conventional $0.05$ significance threshold, so the difference is statistically insignificant. This demonstrates that outbursts do not induce a measurable phase shift at the local level.

Consistent with the local analysis, we performed a model-dependent residual analysis by applying our global orbital modeling procedure (\textit{LTT34+Quad, e=0}) to the full dataset. 
When the best-fit model is subtracted, the median residuals of the active and quiescent data points show a negligible difference of only $\sim 0.6$ seconds. Once again, this yields no statistical significance ($p \approx 0.81 > 0.05$).
Ultimately, both the local and model-dependent statistical tests confirm that active states in HT Cas do not result in systematic long-term timing deviations. Thus, selectively discarding valuable outburst timings is mathematically and observationally unjustified.

\section{Energetic Budget of Magnetic Activity Cycles}
\label{sec:4}
As an alternative to the presence of circumbinary companions, cyclical variations observed in the $O-C$ diagrams of close binaries can arise from magnetic activity cycles in the late-type secondary star. This phenomenon, known as the Applegate mechanism \citep{1992ApJ...385..621A}, suggests that solar-like magnetic cycles can redistribute angular momentum within the star, altering its quadrupole moment and consequently inducing periodic orbital period modulations.

To test whether the complex $O-C$ variations of HT Cas could be driven by such non-planetary mechanisms, we calculated the required energy budget, $\Delta E$, as a fraction of the total available energy in the magnetically active secondary star, $E_{sec}$. We evaluated these energy ratios for both the outer ($LTT_3$) and inner ($LTT_4$) modulations obtained from three commonly used analytical frameworks:
\begin{enumerate}
    \item The thin-shell approximation re-formulated by \citet{2009Ap&SS.319..119T}, which integrates standard Applegate's original concept with the improved stellar quadrupole moment variations model of \citet{1998MNRAS.296..893L}.
    \item The \textit{Two-zone} finite-shell model developed by \citet{2016A&A...587A..34V}, which incorporates distinct and more realistic density profiles for the stellar core and shell ($\Delta E/E_{sec}^-$).
    \item The \textit{Spin-orbit coupling} model introduced by \citet{2020MNRAS.491.1820L}, in which the orbital modulation is driven by the coupling of the active star's spin with the orbital motion via a non-axisymmetric quadrupole moment ($\Delta E_{rot}/E_{2}$).
\end{enumerate}

In all Applegate-like models, the absolute theoretical limit for the magnetic mechanism is $\Delta E/E_{sec} < 1$. However, for the mechanism to be physically sustainable without inducing unobserved stellar luminosity variations, the required energy should ideally be a small fraction of the available budget (i.e., $\Delta E/E_{sec} \ll 1$). Values approaching or significantly exceeding unity imply that the secondary star lacks the energy required to drive the observed $O-C$ amplitude, ruling out a magnetic origin.

To perform these energy budget calculations, we adopted the fundamental astrophysical parameters of the HT Cas system as $M_{sec}=0.09$ $M_\odot$, $R_{sec}=0.154$ $R_\odot$, and $a_{bin}=0.658$ $R_\odot$ derived from comprehensive spectroscopic and photometric studies in the literature \citep{1990ApJ...357..621M, 1991ApJ...378..271H}. The effective temperature of the fully convective secondary was taken as $T_{sec}\approx2500$ $K$, which yields a quiescent luminosity of $L_{sec}\approx0.00083$ $L_\odot$ derived via the Stefan-Boltzmann law \citep[as calculated by][]{2023ApJ...953...63H}. Furthermore, for the \textit{Two-zone} model calculations, we adopted the structural constants $k_1=0.133$ and $k_2=3.42$ as suggested by \citet{2016A&A...587A..34V} for the fully convective low-mass secondary stars. The calculated energy ratios for our proposed two-companion models are summarized at the bottom of Table \ref{tab:htcas_models}.

Under the simplistic Standard model \citep{2009Ap&SS.319..119T}, the required energy ratios for both the outer and inner modulations remain mathematically below the absolute limit ($\Delta E/E_{sec} \approx 0.04 - 0.10$). However, the thin-shell assumption fundamentally contradicts the internal structure of late-type secondary stars in CVs like HT Cas, which are fully convective. Thus, this approximation underestimates the true energy budget required to redistribute angular momentum in such deeply convective stars.

When applying the more comprehensive physical frameworks that account for realistic stellar density profiles, the magnetic hypothesis fails. In our most plausible circular orbit model ($e=0$), the Two-zone model \citep{2016A&A...587A..34V} reveals that the energy required to generate the $LTT_3$ and $LTT_4$ signals exceeds the maximum available stellar energy by factors of 1.72 and 1.60, respectively. Furthermore, under the Spin-orbit coupling framework \citep{2020MNRAS.491.1820L}, the energy constraints fail, demanding approximately 24 and 34 times more energy than the secondary star can physically provide. Even if we consider the unconstrained free-eccentricity model (\textit{LTT34+Quad}), the Spin-orbit model still yields massively prohibitive ratios ($\Delta E_{rot}/E_{sec} > 7$), and the Two-zone model yields values just below unity ($\Delta E/E_{sec}^- \sim 0.68$ and $0.94$). While the latter might technically be $<1$, it violates the condition $\Delta E \ll E_{sec}$ for a sustainable mechanism. 
Therefore, current energetic constraints based on classical frameworks demonstrate that magnetic activity is insufficient to independently drive the observed $O-C$ variations in HT Cas.

\section{Dynamical Stability of Orbital Configurations}
\label{sec:5}
To investigate the long-term dynamical orbital stability of the proposed multi-component models for HT Cas, we utilized the N-body orbital integration package \textit{REBOUND}\footnote{https://rebound.readthedocs.io} \citep{2012A&A...537A.128R}. The dynamical behavior of celestial bodies in a purely gravitational framework was simulated using the Wisdom-Holman symplectic integrator \citep[WHFast,][]{2015MNRAS.452..376R}. WHFast is exceptionally efficient for planetary systems, as it conserves energy by splitting each integration timestep into Keplerian and interaction phases. In our study, we extracted two fundamental diagnostic tools from \textit{REBOUND}: (1) the Mean Exponential Growth factor of Nearby Orbits \citep[MEGNO,][]{2000A&AS..147..205C, 2015MNRAS.452..376R} chaos indicator, which assigns a numerical value ($\langle Y \rangle$) to classify an orbit as regular or chaotic. An orbital configuration is considered dynamically stable if $\langle Y \rangle \le 2$, while values greater than 2 indicate chaotic interactions that eventually lead to orbital destabilization. A value of $\langle Y \rangle \ge 10$ indicates the ejection or collision of a companion in the system. (2) The orbital stability timeline, which tracks the variations in essential orbital parameters (such as the semi-major axis, $a$, and eccentricity, $e$) as a function of time, revealing how mutual gravitational interactions evolve on secular timescales \citep{2021MNRAS.506.2122B, 2023MNRAS.526.4725O}.

Consistent with previous dynamical studies of circumbinary planets \citep[e.g.,][]{1999AJ....117..621H, 2001A&A...378..569G}, the central binary star of HT Cas was treated as a single central mass ($M_{bin} = 0.70 \ M_\odot$). To maximize interaction probabilities and test the system under the most extreme perturbative conditions, all planetary orbits were assumed to be coplanar. The integration timestep was conservatively set to $\sim 0.1$ years (which is strictly less than 1\% of the shortest planetary orbital period in the system) to maintain the symplectic nature of the WHFast algorithm, and the ejection boundary was defined as 100 AU.

We first examined the unconstrained two-companion model (\textit{LTT34+Quad}), where the initial MCMC analysis favored a highly eccentric orbit ($e_3 = 0.94$) for the massive outer companion ($LTT_3$). The N-body integration for this configuration revealed an extremely chaotic dynamical environment. As prominently depicted in the orbital stability timeline, the massive third body ($53.5 \ M_{Jup}$) combined with its high eccentric trajectory induces strong mutual gravitational perturbations. These perturbations catastrophically disrupt the system's configuration almost instantly. Within a remarkably short timescale of just $\sim 800$ years, the eccentricity of the inner companion ($LTT_4$) is rapidly forced to unity, accompanied by an unbounded divergence in its semi-major axis. This behavior indicates ejection of the inner body from the system. Consequently, the MEGNO values for both bodies immediately reached $\langle Y \rangle \ge 10$, confirming that such a high-eccentricity configuration is dynamically unstable and physically impossible to sustain.

\begin{figure}[ht!]
    \centering
    \includegraphics[width=0.9\columnwidth]{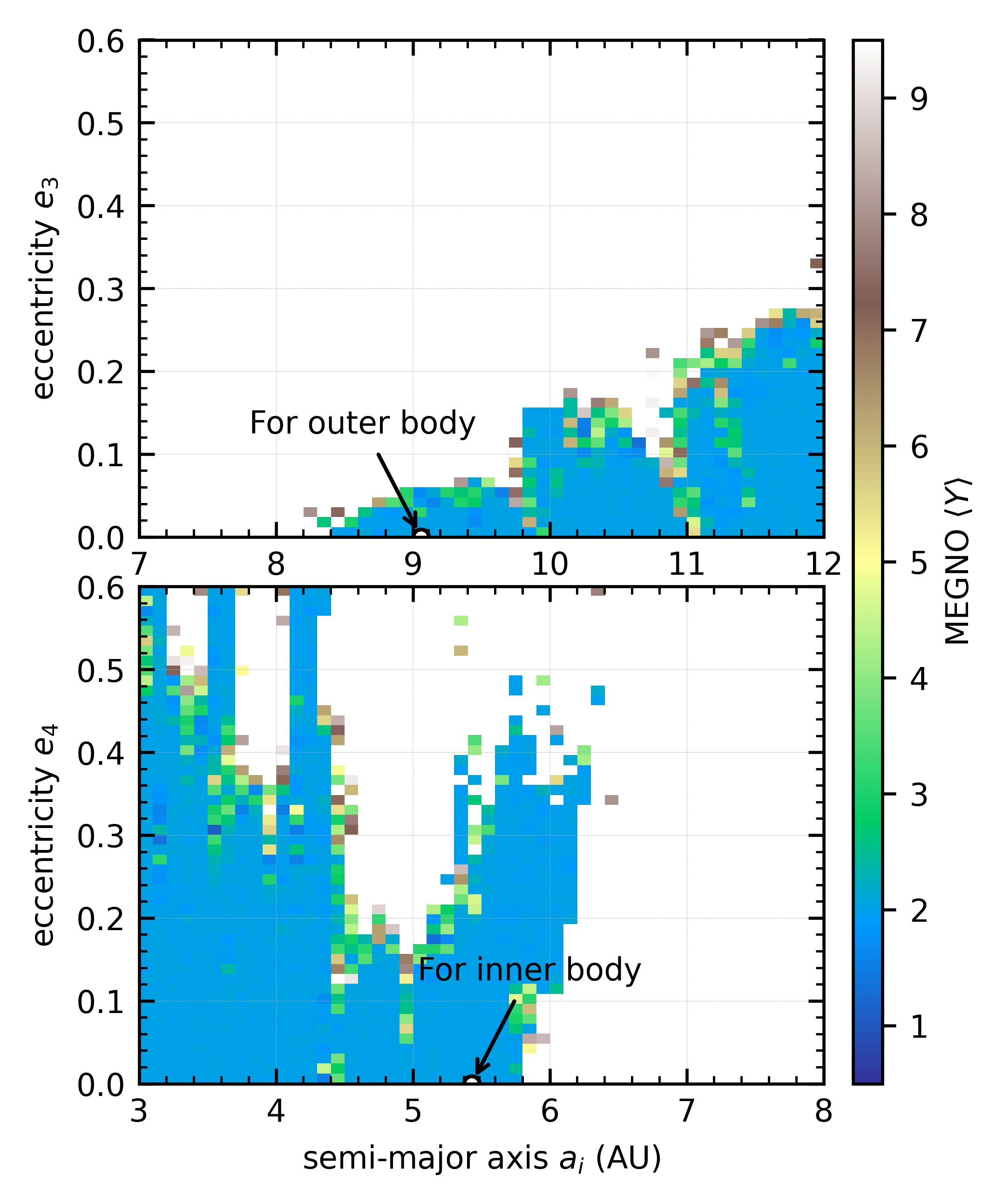}
    \caption{Map of the MEGNO chaos indicator ($\langle Y \rangle$) for the dynamically stable circular two-companion model (\textit{LTT34+Quad, $e=0$}) integrated over $10^6$ years. The parameter space was explored by systematically varying the semi-major axis and eccentricity of the outer companion ($LTT_3$, top panel) and the inner companion ($LTT_4$, bottom panel), while keeping the non-varying body fixed at its nominal circular values. The white circles denote the best-fit orbital solutions from the MCMC analysis (Table \ref{tab:htcas_models}).}
    \label{fig:stability_map}
\end{figure}

Subsequently, we tested the circular two-companion model (\textit{LTT34+Quad, $e=0$}), using the corresponding orbital parameters from Table \ref{tab:htcas_models}. For this configuration, we mapped the MEGNO parameter surface over an integration time of $10^6$ years by separately varying the semi-major axis and eccentricity of each companion while keeping the other fixed at its nominal values. 
The resulting MEGNO stability map is presented in Figure \ref{fig:stability_map}. The stability analysis reveals that the nominal orbital solutions for both the outer and inner bodies lie precisely within the stable region ($\langle Y \rangle \le 2$). The maps clearly demonstrate that any significant deviation from the circular assumption directly drives the system into the chaotic, unstable regime ($\langle Y \rangle > 2$). To further validate this configuration, we ran an orbital stability timeline integration for $10^7$ years. Throughout this extended timescale, the circular model showed no signs of orbital crossing or chaotic divergence. Both companions exhibited only minor secular oscillations in their semi-major axes and maintained their eccentricities at near-zero values, demonstrating a stable and long-lasting dynamical hierarchy.

\section{Discussion and Conclusions}
\label{sec:discussion}

Using an extended and precise observational baseline spanning from 1978 to 2026, we presented a comprehensive analysis of the long-term orbital period variations in the short-period eclipsing dwarf nova HT Cas. Our analysis integrated mid-eclipse timings from both new ground-based observations and extensive archival databases. A model-independent statistical analysis of these timings demonstrated that outbursts in HT Cas do not introduce systematic, long-term phase shifts ($\sim3.6$ s, $p \approx 0.067$). This finding is further supported by our model-dependent residual analysis, which also yielded a negligible median offset ($\sim 0.6$ s, $p \approx 0.81$). This lack of systematic offset implies that the accretion disk asymmetry, typically produced by a dominant bright spot, is weak in HT Cas. A similar behavior was reported for IR Com, a photometric twin of HT Cas, where outburst and quiescence timings showed no significant discrepancy \citep{2002PASJ...54...79K}. 
Consequently, selectively discarding valuable outburst timings is mathematically and observationally unjustified, ensuring that our $\sim$48-year dataset provides a robust and unbiased foundation for evaluating the secular and cyclic dynamics of the system.

A significant outcome of our multi-body modeling is the resolution of the secular evolution anomaly of HT Cas. An initial, simple quadratic fit to the $O-C$ data yielded a statistically significant positive quadratic term ($Q = +1.846 \times 10^{-14}$ days), which would imply a secular orbital expansion ($\dot{P} \simeq 5.01 \times 10^{-13}$). However, this directly contradicts the standard evolutionary theory of short-period CVs below the period gap, where AML—driven primarily by gravitational radiation—should strictly force the orbit to shrink \citep[$Q < 0$,][]{2011ApJS..194...28K}. Our analysis demonstrates that this apparent positive period derivative is not a true secular evolution, but rather a geometric artifact of fitting a simple parabola to a segment of a much longer, incomplete cyclic modulation. Indeed, when the gravitational perturbations (LTT effects) of the proposed circumbinary companions are accounted for in the MCMC models, this positive trend is completely rectified. The quadratic term reverts to a physically expected negative value ($Q = -1.23^{+0.60}_{-0.54} \times 10^{-14}$ days). Using the relation $\dot{P} = 2Q/P_{bin}$, this corrected term yields a secular orbital decay rate of $\dot{P} \approx -3.34 \times 10^{-13}$. While standard gravitational radiation alone predicts a slower rate of period decrease ($\dot{P}_{GR} \approx -1.03 \times 10^{-13}$) for the known parameters of HT Cas \citep{2023ApJ...953...63H}, our derived value aligns with the necessity of additional AML mechanisms to explain the observed orbital decay in this system  \citep[e.g.][]{ 2023ApJ...953...63H}. Thus, this quadratic term reinforces the physical necessity and reliability of our multi-periodic LTT architecture.

Unlike previous studies that relied on single-companion approximations \citep{2008A&A...480..481B, 2023ApJ...953...63H}, our analysis establishes that the long-term $O-C$ variability of HT Cas can only be fully resolved by adopting a hierarchical two-companion configuration. The initial free-eccentricity model (\textit{LTT34+Quad}) identified two companions with highly eccentric orbits ($e_3 \sim 0.94$, $e_4 \sim 0.51$) and resulted in a massive outer companion of $\sim 53.5\, M_{Jup}$. However, as demonstrated by our MCMC posterior distributions, this high-eccentricity result is a mathematical consequence of the algorithm favoring extreme geometric parameters ($e$ and $K$) to forcefully overfit sharp $O-C$ variations. These sharp variations are a natural consequence of sparse data coverage (observational gaps) and the photometric flickering inherent to the accretion dynamics of CVs, which is robustly captured by our systematic uncertainty parameter ($\sigma_f \sim 7.2$ s). By efficiently absorbing this background activity noise, the systematic uncertainty term shows that these sharp peaks do not require high orbital eccentricities.

N-body dynamical simulations confirmed that these extreme parameters are unsustainable; the highly eccentric unconstrained model collapsed almost instantly, ejecting the inner companion within $\sim 800$ years and resulting in highly chaotic MEGNO values ($\langle Y \rangle \ge 10$). This result is consistent with the literature, which demonstrates that hosting multiple massive companions with such high eccentricities inevitably triggers strong secular interactions and chaotic orbital evolution, rendering these configurations dynamically implausible \citep[e.g.,][]{2010MNRAS.407.2362P, 2011MNRAS.416L..11H, 2012MNRAS.427.2812H, 2012AJ....144...34H, 2013MNRAS.435.2033H, 2025NewA..11902414E}.
The resulting circular two-companion model (\textit{LTT34+Quad, e=0}) constrains the orbits to $P_3 = 32.6$ and $P_4 = 15.1$ years, yielding physically realistic minimum masses of $M_3 \sin i_3 \sim 9.8\, M_{Jup}$ and $M_4 \sin i_4 \sim 5.0\, M_{Jup}$ at distances of $\sim 9.06$ AU and $\sim 5.45$ AU, respectively. The dynamical analysis provides compelling physical justification for rejecting the spurious high-eccentricity solution and validates the circular, coplanar two-companion configuration, which maintained a stable dynamical configuration ($\langle Y \rangle \le 2$) over a $10^7$-year integration timeline. 
It is noteworthy that the secular AML of the central binary does not undermine this dynamical stability. Given our derived orbital decay rate ($\dot{P} \approx -3.34 \times 10^{-13}$), the characteristic evolutionary timescale of the system, defined as $\tau = P_{bin} / |\dot{P}|$ \citep{2011ApJS..194...28K}, is approximately $6 \times 10^8$ years. Because this timescale is more than an order of magnitude longer than our $10^7$-year N-body integration, the circumbinary planetary orbits will adjust adiabatically to the slowly shrinking central binary. Therefore, secular AML can be safely neglected when assessing the resonant stability and chaotic interactions of the proposed companions over million-year timescales.

We also evaluated the Applegate mechanism as an alternative explanation for the cyclic modulations. While fundamental magnetic cycles may inherently exist in the fully convective low-mass secondary star ($0.09 M_\odot$) of HT Cas, our energetic budget evaluations using finite-shell and spin-orbit coupling frameworks \citep{2016A&A...587A..34V, 2020MNRAS.491.1820L} showed that the energy required to drive the observed $O-C$ amplitudes exceeds the secondary star's available luminosity by factors ranging from 1.6 to 34. This severe energy deficit demonstrates that the magnetic mechanism alone cannot independently drive the two observed sinusoidal variations.
However, recent population studies of white dwarf binaries \citep{2016MNRAS.460.3873B, 2026MNRAS.547ag358Y} suggest ruling out magnetic activity solely based on this classical energy deficit may be premature. These studies demonstrate that the vast majority of WD+dM binaries fail this exact energy test, yet the amplitude of their eclipse timing variations strongly correlates with the secondary star's parameters, heavily implying a fundamental magnetic origin. Newly advanced theoretical frameworks \citep[such as Azimuthal Dynamo Waves;][]{2026arXiv260427609N}, which can generate large-amplitude variations, may resolve this ubiquitous energy crisis without violating classical energetic limits. Consequently, classical Applegate-type models require further theoretical development to accurately describe the magnetic mechanisms operating in these binaries.

Our comprehensive $\sim$48-year baseline successfully covers approximately 1.5 orbital cycles of the outermost body ($P_3 \sim 32.6$ years). While capturing three critical extrema mathematically constrains the sinusoidal parameters, definitively confirming long-term periodic repeatability naturally requires longer observational baselines. To ensure our model does not overfit, an LS periodogram analysis of the $O-C$ residuals confirmed the absence of any significant peaks above the FAP threshold, indicating that the residuals are consistent with white noise.

Predicting future eclipse times in PCEB systems remains highly challenging \citep[e.g.,][]{2025MNRAS.544...24P}. The literature contains numerous instances where proposed LTT models eventually failed to predict future eclipse arrival times as new data were acquired. For instance, the most recent LTT model proposed for HT Cas by \citet{2023ApJ...953...63H} is now in complete disagreement with the current $O-C$ distribution, diverging sharply downwards. 
Consequently, recent statistical studies \citep[e.g.,][]{2024ApJ...972...33S} that rely on such outdated models may reach inaccurate conclusions regarding the underlying physical mechanisms. The deviation of a previous model emphasizes the necessity of iterative parameter refinement.

On the other hand, a circumbinary planetary system remains a physically robust scenario. Recently, a statistical study by \citet{2024ApJ...972...33S} questioned the circumbinary planet hypothesis in several CVs, including HT Cas, arguing that the third-body hypothesis is statistically inconsistent, partially due to an observed correlation suggesting mass loss from the third-body over orbital time. However, this conclusion relies on the assumption that these are "first-generation" planets that survived the violent common-envelope (CE) phase. Survival of Jovian-mass planets during CE evolution is dynamically and energetically improbable \citep{2010MNRAS.408..631N}. In a PCEB like HT Cas, the planetary companions are more likely to be "second-generation" planets formed from the remnant circumbinary gas and dust disk expelled during the CE phase \citep{2013A&A...549A..95Z, 2015AN....336..458S, 2016RSOS....350571V, 2025ApJ...994...67E}. In this context, the planetary masses do not represent a secular mass loss, but rather reflect the initial conditions and mass budget of the ejected CE material. Furthermore, \citet{2024ApJ...972...33S} evaluated HT Cas assuming a single-companion model with $P \sim 36$ years, which fails to capture the multi-periodic nature of the system we have demonstrated here. When the system is properly modeled as a hierarchical two-companion architecture, the parameters are consistent with dynamically stable, second-generation circumbinary planet formation scenarios. 
Unlike many other systems where proposed planetary fits ultimately fail rigorous dynamical stability criteria \citep{2026MNRAS.547ag358Y}, our two-companion circular model maintains complete dynamical stability over a $10^7$-year N-body integration.

We also note that the lack of prominent astrometric excess noise in Gaia DR3 does not contradict the presence of these proposed companions. Because the orbital period of the outermost companion ($P_3 \sim 32.6$ years) significantly exceeds the Gaia DR3 observational baseline ($\sim 34$ months), the expected astrometric displacement traces a nearly linear path. As reported by recent studies \citep{2023MNRAS.526.2241B, 2026ApJ...998..155X}, the Gaia data processing pipeline absorbs such linear motions directly into the proper motion vector, effectively hiding the companion's astrometric signature within the intrinsic noise limit. Consequently, future astrometric releases with extended full-epoch baselines, such as Gaia DR4 or subsequent catalogs, will be required to overcome this proper motion absorption and reveal the expected non-linear transverse components of the astrometric wobble \citep{2024A&A...691A.115Z}.

In conclusion, based on our most plausible circular model, the $\sim$48-year data baseline covers more than one full orbital cycle of the outermost body. 
Our proposed hierarchical two-companion model stands as a highly robust physical framework that successfully survives both rigorous N-body dynamical stability tests and the energetic limitations of classical Applegate mechanisms.
Given that both a dynamically stable planetary architecture and advanced magnetic models remain physically plausible explanations, continued high-precision photometric monitoring and upcoming extended-baseline astrometric missions will be crucial. These future observations are required to map the finer details of the turnaround phases, test for slight non-zero orbital eccentricities, and ultimately distinguish between complex magnetic cycles and circumbinary companions.

\begin{acknowledgments}
In this study, observational data obtained within the scope of the project numbered T100-631 and T100-1333, conducted using the TUG100 Telescope (Antalya) at the site under the Türkiye National Observatories, have been utilized, and we express our gratitude for the invaluable support provided by the Türkiye National Observatories, the observation team, and all its staff. We also wish to thank Adıyaman University Astrophysics Application and Research Center (Türkiye) for the allocation of their telescope time.
We gratefully acknowledge the use of the mid-eclipse timing data provided by the Variable Star and Exoplanet Section of the Czech Astronomical Society, available through the VarAstro (O–C Gateway) database. We gratefully acknowledge the contributions of the AAVSO observer community, whose photometric data and metadata resources were used in this study and made available through the AAVSO’s scientific archives. This work includes data obtained through the Scientific and Technological Research Council of Turkey (TUBITAK), under project number 114F460 (I. Nasiroglu and H. Er). A. Takey and M. Abdelkareem acknowledge financial support from the Egyptian Science, Technology \& Innovation Funding Authority (STDF) under grant number 48102. A. Ozdonmez acknowledges financial support from the BAGEP Award of the Science Academy. I. Nasiroglu and B.B. Gürbulak acknowledge support from Atatürk University through BAP Project FDK-2024-14239. Furthermore, B.B. Gürbulak acknowledges support from the TÜBİTAK Scientist Support Programs Presidency (BİDEB) under the 2211-A National PhD Scholarship Program. 

\end{acknowledgments}

\facility{AAVSO (AID)}

\section*{Data Availability Statement}
The data underlying this article, specifically the mid-eclipse times presented in Table \ref{tab:midtimes_htcas}, are available in a machine-readable format in the online supplementary material.






\bibliography{sample701}{}

@ARTICLE{2022ApJ...935..167A,
       author = {{Astropy Collaboration} and {Price-Whelan}, Adrian M. and {Lim}, Pey Lian and {Earl}, Nicholas and {Starkman}, Nathaniel and {Bradley}, Larry and {Shupe}, David L. and {Patil}, Aarya A. and {Corrales}, Lia and {Brasseur}, C.~E. and {N{\"o}the}, Maximilian and {Donath}, Axel and {Tollerud}, Erik and {Morris}, Brett M. and {Ginsburg}, Adam and {Vaher}, Eero and {Weaver}, Benjamin A. and {Tocknell}, James and {Jamieson}, William and {van Kerkwijk}, Marten H. and {Robitaille}, Thomas P. and {Merry}, Bruce and {Bachetti}, Matteo and {G{\"u}nther}, H. Moritz and {Aldcroft}, Thomas L. and {Alvarado-Montes}, Jaime A. and {Archibald}, Anne M. and {B{\'o}di}, Attila and {Bapat}, Shreyas and {Barentsen}, Geert and {Baz{\'a}n}, Juanjo and {Biswas}, Manish and {Boquien}, M{\'e}d{\'e}ric and {Burke}, D.~J. and {Cara}, Daria and {Cara}, Mihai and {Conroy}, Kyle E. and {Conseil}, Simon and {Craig}, Matthew W. and {Cross}, Robert M. and {Cruz}, Kelle L. and {D'Eugenio}, Francesco and {Dencheva}, Nadia and {Devillepoix}, Hadrien A.~R. and {Dietrich}, J{\"o}rg P. and {Eigenbrot}, Arthur Davis and {Erben}, Thomas and {Ferreira}, Leonardo and {Foreman-Mackey}, Daniel and {Fox}, Ryan and {Freij}, Nabil and {Garg}, Suyog and {Geda}, Robel and {Glattly}, Lauren and {Gondhalekar}, Yash and {Gordon}, Karl D. and {Grant}, David and {Greenfield}, Perry and {Groener}, Austen M. and {Guest}, Steve and {Gurovich}, Sebastian and {Handberg}, Rasmus and {Hart}, Akeem and {Hatfield-Dodds}, Zac and {Homeier}, Derek and {Hosseinzadeh}, Griffin and {Jenness}, Tim and {Jones}, Craig K. and {Joseph}, Prajwel and {Kalmbach}, J. Bryce and {Karamehmetoglu}, Emir and {Ka{\l}uszy{\'n}ski}, Miko{\l}aj and {Kelley}, Michael S.~P. and {Kern}, Nicholas and {Kerzendorf}, Wolfgang E. and {Koch}, Eric W. and {Kulumani}, Shankar and {Lee}, Antony and {Ly}, Chun and {Ma}, Zhiyuan and {MacBride}, Conor and {Maljaars}, Jakob M. and {Muna}, Demitri and {Murphy}, N.~A. and {Norman}, Henrik and {O'Steen}, Richard and {Oman}, Kyle A. and {Pacifici}, Camilla and {Pascual}, Sergio and {Pascual-Granado}, J. and {Patil}, Rohit R. and {Perren}, Gabriel I. and {Pickering}, Timothy E. and {Rastogi}, Tanuj and {Roulston}, Benjamin R. and {Ryan}, Daniel F. and {Rykoff}, Eli S. and {Sabater}, Jose and {Sakurikar}, Parikshit and {Salgado}, Jes{\'u}s and {Sanghi}, Aniket and {Saunders}, Nicholas and {Savchenko}, Volodymyr and {Schwardt}, Ludwig and {Seifert-Eckert}, Michael and {Shih}, Albert Y. and {Jain}, Anany Shrey and {Shukla}, Gyanendra and {Sick}, Jonathan and {Simpson}, Chris and {Singanamalla}, Sudheesh and {Singer}, Leo P. and {Singhal}, Jaladh and {Sinha}, Manodeep and {Sip{\H{o}}cz}, Brigitta M. and {Spitler}, Lee R. and {Stansby}, David and {Streicher}, Ole and {{\v{S}}umak}, Jani and {Swinbank}, John D. and {Taranu}, Dan S. and {Tewary}, Nikita and {Tremblay}, Grant R. and {de Val-Borro}, Miguel and {Van Kooten}, Samuel J. and {Vasovi{\'c}}, Zlatan and {Verma}, Shresth and {de Miranda Cardoso}, Jos{\'e} Vin{\'\i}cius and {Williams}, Peter K.~G. and {Wilson}, Tom J. and {Winkel}, Benjamin and {Wood-Vasey}, W.~M. and {Xue}, Rui and {Yoachim}, Peter and {Zhang}, Chen and {Zonca}, Andrea and {Astropy Project Contributors}},
        title = "{The Astropy Project: Sustaining and Growing a Community-oriented Open-source Project and the Latest Major Release (v5.0) of the Core Package}",
      journal = {\apj},
         year = 2022,
        month = Aug,
       volume = {935},
       number = {2},
          eid = {167},
        pages = {167},
          doi = {10.3847/1538-4357/ac7c74},
archivePrefix = {arXiv},
       eprint = {2206.14220},
 primaryClass = {astro-ph.IM},
       adsurl = {https://ui.adsabs.harvard.edu/abs/2022ApJ...935..167A}
}

@ARTICLE{2011ApJS..194...28K,
       author = {{Knigge}, Christian and {Baraffe}, Isabelle and {Patterson}, Joseph},
        title = "{The Evolution of Cataclysmic Variables as Revealed by Their Donor Stars}",
      journal = {\apjs},
         year = 2011,
        month = jun,
       volume = {194},
       number = {2},
          eid = {28},
        pages = {28},
          doi = {10.1088/0067-0049/194/2/28},
archivePrefix = {arXiv},
       eprint = {1102.2440},
 primaryClass = {astro-ph.SR},
       adsurl = {https://ui.adsabs.harvard.edu/abs/2011ApJS..194...28K}
}

@ARTICLE{1981A&A...100L...7V,
       author = {{Verbunt}, F. and {Zwaan}, C.},
        title = "{Magnetic braking in low-mass X-ray binaries.}",
      journal = {\aap},
         year = 1981,
        month = jul,
       volume = {100},
        pages = {L7-L9},
       adsurl = {https://ui.adsabs.harvard.edu/abs/1981A&A...100L...7V}
}

@ARTICLE{1962ApJ...136..312K,
       author = {{Kraft}, Robert P. and {Mathews}, Jon and {Greenstein}, Jesse L.},
        title = "{Binary Stars among Cataclysmic Variables. II. Nova WZ Sagittae: a Possible Radiator of Gravitational Waves.}",
      journal = {\apj},
         year = 1962,
        month = jul,
       volume = {136},
        pages = {312-315},
          doi = {10.1086/147381},
       adsurl = {https://ui.adsabs.harvard.edu/abs/1962ApJ...136..312K}
}

@ARTICLE{2023ApJ...953...63H,
       author = {{Han}, Z.-T. and {Qian}, S.-B. and {Han}, Q.-W. and {Zang}, L. and {Soonthornthum}, B. and {Li}, L.-J. and {Zhu}, L.-Y. and {Liu}, W. and {Fern{\'a}ndez Laj{\'u}s}, E. and {Dai}, Z.-B. and {Na}, W.-W.},
        title = "{Orbital Period Variations in HT Cas: Evidence for Additional Angular Momentum Loss and a High-eccentricity Giant Planet}",
      journal = {\apj},
         year = 2023,
        month = aug,
       volume = {953},
       number = {1},
          eid = {63},
        pages = {63},
          doi = {10.3847/1538-4357/acdd6e},
       adsurl = {https://ui.adsabs.harvard.edu/abs/2023ApJ...953...63H}
}

@ARTICLE{2020ApJ...901..113F,
       author = {{Fang}, Xiaohui and {Qian}, Shengbang and {Han}, Zhongtao and {Wang}, Qishan},
        title = "{Long-term Period Changes and Brightness Variations for the Deeply Eclipsing Cataclysmic Variable SW Sex}",
      journal = {\apj},
         year = 2020,
        month = oct,
       volume = {901},
       number = {2},
          eid = {113},
        pages = {113},
          doi = {10.3847/1538-4357/abb1b9},
       adsurl = {https://ui.adsabs.harvard.edu/abs/2020ApJ...901..113F}
}

@ARTICLE{2022NewA...9301751S,
       author = {{Sun}, Qi-Bin and {Qian}, Sheng-Bang and {Dong}, Ai-Jun and {Zhi}, Qi-Jun and {Han}, Zhong-Tao and {Liu}, Wei and {Chang}, Xin and {Liu}, Chang and {Xiang}, Hong-Bin and {Peng}, Xue-Bing and {Zhang}, Bin and {Zhang}, Xu-Dong and {Fern{\'a}ndez Laj{\'u}s}, E.},
        title = "{Study on the variation of orbital period, quasi-periodic oscillations and negative superhumps in V729 Sgr}",
      journal = {\na},
         year = 2022,
        month = may,
       volume = {93},
          eid = {101751},
        pages = {101751},
          doi = {10.1016/j.newast.2021.101751},
       adsurl = {https://ui.adsabs.harvard.edu/abs/2022NewA...9301751S}
}

@ARTICLE{1992ApJ...385..621A,
       author = {{Applegate}, James H.},
        title = "{A Mechanism for Orbital Period Modulation in Close Binaries}",
      journal = {\apj},
         year = 1992,
        month = feb,
       volume = {385},
        pages = {621},
          doi = {10.1086/170967},
       adsurl = {https://ui.adsabs.harvard.edu/abs/1992ApJ...385..621A}
}

@ARTICLE{1998MNRAS.296..893L,
       author = {{Lanza}, A.~F. and {Rodono}, M. and {Rosner}, R.},
        title = "{Orbital period modulation and magnetic cycles in close binaries}",
      journal = {\mnras},
         year = 1998,
        month = jun,
       volume = {296},
       number = {4},
        pages = {893-902},
          doi = {10.1046/j.1365-8711.1998.01446.x},
       adsurl = {https://ui.adsabs.harvard.edu/abs/1998MNRAS.296..893L}
}

@ARTICLE{2011A&A...526A..53B,
       author = {{Beuermann}, K. and {Buhlmann}, J. and {Diese}, J. and {Dreizler}, S. and {Hessman}, F.~V. and {Husser}, T.-O. and {Miller}, G.~F. and {Nickol}, N. and {Pons}, R. and {Ruhr}, D. and {Schm{\"u}lling}, H. and {Schwope}, A.~D. and {Sorge}, T. and {Ulrichs}, L. and {Winget}, D.~E. and {Winget}, K.~I.},
        title = "{The giant planet orbiting the cataclysmic binary DP Leonis}",
      journal = {\aap},
         year = 2011,
        month = feb,
       volume = {526},
          eid = {A53},
        pages = {A53},
          doi = {10.1051/0004-6361/201015942},
archivePrefix = {arXiv},
       eprint = {1011.3905},
 primaryClass = {astro-ph.SR},
       adsurl = {https://ui.adsabs.harvard.edu/abs/2011A&A...526A..53B}
}

@ARTICLE{2015MNRAS.448.1118G,
       author = {{Go{\'z}dziewski}, K. and {S{\l}owikowska}, A. and {Dimitrov}, D. and {Krzeszowski}, K. and {{\.Z}ejmo}, M. and {Kanbach}, G. and {Burwitz}, V. and {Rau}, A. and {Irawati}, P. and {Richichi}, A. and {Gawro{\'n}ski}, M. and {Nowak}, G. and {Nasiroglu}, I. and {Kubicki}, D.},
        title = "{The HU Aqr planetary system hypothesis revisited}",
      journal = {\mnras},
         year = 2015,
        month = apr,
       volume = {448},
       number = {2},
        pages = {1118-1136},
          doi = {10.1093/mnras/stu2728},
archivePrefix = {arXiv},
       eprint = {1412.5899},
 primaryClass = {astro-ph.EP},
       adsurl = {https://ui.adsabs.harvard.edu/abs/2015MNRAS.448.1118G}
}

@ARTICLE{2025NewA..11902414E,
       author = {{Er}, Huseyin},
        title = "{The dynamical orbital stability of the proposed <mml:math><mml:mrow><mml:mn>2</mml:mn><mml:mo>+</mml:mo><mml:mn>1</mml:mn><mml:mo>+</mml:mo><mml:mn>1</mml:mn></mml:mrow></mml:math> hierarchical eclipsing binary systems}",
      journal = {\na},
         year = 2025,
        month = oct,
       volume = {119},
          eid = {102414},
        pages = {102414},
          doi = {10.1016/j.newast.2025.102414},
       adsurl = {https://ui.adsabs.harvard.edu/abs/2025NewA..11902414E}
}

@ARTICLE{2010MNRAS.404..837H,
       author = {{Hinse}, T.~C. and {Christou}, A.~A. and {Alvarellos}, J.~L.~A. and {Go{\'z}dziewski}, K.},
        title = "{Application of the MEGNO technique to the dynamics of Jovian irregular satellites}",
      journal = {\mnras},
         year = 2010,
        month = may,
       volume = {404},
       number = {2},
        pages = {837-857},
          doi = {10.1111/j.1365-2966.2010.16307.x},
archivePrefix = {arXiv},
       eprint = {0907.4886},
 primaryClass = {astro-ph.EP},
       adsurl = {https://ui.adsabs.harvard.edu/abs/2010MNRAS.404..837H}
}

@ARTICLE{2012MNRAS.425..930G,
       author = {{Go{\'z}dziewski}, Krzysztof and {Nasiroglu}, Ilham and {S{\l}owikowska}, Aga and {Beuermann}, Klaus and {Kanbach}, Gottfried and {Gauza}, Bartosz and {Maciejewski}, Andrzej J. and {Schwarz}, Robert and {Schwope}, Axel D. and {Hinse}, Tobias C. and {Haghighipour}, Nader and {Burwitz}, Vadim and {S{\l}onina}, Mariusz and {Rau}, Arne},
        title = "{On the HU Aquarii planetary system hypothesis}",
      journal = {\mnras},
         year = 2012,
        month = sep,
       volume = {425},
       number = {2},
        pages = {930-949},
          doi = {10.1111/j.1365-2966.2012.21341.x},
archivePrefix = {arXiv},
       eprint = {1205.4164},
 primaryClass = {astro-ph.EP},
       adsurl = {https://ui.adsabs.harvard.edu/abs/2012MNRAS.425..930G}
}

@ARTICLE{2011MNRAS.416L..11H,
       author = {{Horner}, J. and {Marshall}, J.~P. and {Wittenmyer}, Robert A. and {Tinney}, C.~G.},
        title = "{A dynamical analysis of the proposed HU Aquarii planetary system}",
      journal = {\mnras},
         year = 2011,
        month = sep,
       volume = {416},
       number = {1},
        pages = {L11-L15},
          doi = {10.1111/j.1745-3933.2011.01087.x},
archivePrefix = {arXiv},
       eprint = {1106.0777},
 primaryClass = {astro-ph.EP},
       adsurl = {https://ui.adsabs.harvard.edu/abs/2011MNRAS.416L..11H}
}

@ARTICLE{2016A&A...587A..34V,
       author = {{V{\"o}lschow}, M. and {Schleicher}, D.~R.~G. and {Perdelwitz}, V. and {Banerjee}, R.},
        title = "{Eclipsing time variations in close binary systems: Planetary hypothesis vs. Applegate mechanism}",
      journal = {\aap},
         year = 2016,
        month = mar,
       volume = {587},
          eid = {A34},
        pages = {A34},
          doi = {10.1051/0004-6361/201527333},
archivePrefix = {arXiv},
       eprint = {1512.01960},
 primaryClass = {astro-ph.SR},
       adsurl = {https://ui.adsabs.harvard.edu/abs/2016A&A...587A..34V}
}

@ARTICLE{1943AN....274...36H,
       author = {{Hoffmeister}, Cuno},
        title = "{213 neue Ver{\"a}nderliche}",
      journal = {Astronomische Nachrichten},
         year = 1943,
        month = jul,
       volume = {274},
        pages = {36},
          doi = {10.1002/asna.19432740109},
       adsurl = {https://ui.adsabs.harvard.edu/abs/1943AN....274...36H}
}

@INPROCEEDINGS{1979BAAS...11..664P,
       author = {{Patterson}, J.},
        title = "{HT Cassiopeiae: Rosetta Stone for Accretion Processes in Dwarf Novae}",
    booktitle = {Bulletin of the American Astronomical Society},
         year = 1979,
       volume = {11},
        month = dec,
        pages = {664},
       adsurl = {https://ui.adsabs.harvard.edu/abs/1979BAAS...11..664P}
}

@ARTICLE{1981ApJS...45..517P,
       author = {{Patterson}, J.},
        title = "{Rapid oscillations in cataclysmic variables. VI. Periodicities in erupting dwarf novae.}",
      journal = {\apjs},
         year = 1981,
        month = mar,
       volume = {45},
        pages = {517-539},
          doi = {10.1086/190723},
       adsurl = {https://ui.adsabs.harvard.edu/abs/1981ApJS...45..517P}
}

@ARTICLE{1991ApJ...378..271H,
       author = {{Horne}, Keith and {Wood}, Janet H. and {Stiening}, Rae F.},
        title = "{Eclipse Studies of the Dwarf Nova HT Cassiopeiae. I. Observations and System Parameters}",
      journal = {\apj},
         year = 1991,
        month = sep,
       volume = {378},
        pages = {271},
          doi = {10.1086/170426},
       adsurl = {https://ui.adsabs.harvard.edu/abs/1991ApJ...378..271H}
}

@ARTICLE{1981ApJ...245.1035Y,
       author = {{Young}, P. and {Schneider}, D.~P. and {Shectman}, S.~A.},
        title = "{The voracious vortex in HT Cassiopeiae.}",
      journal = {\apj},
         year = 1981,
        month = may,
       volume = {245},
        pages = {1035-1042},
          doi = {10.1086/158880},
       adsurl = {https://ui.adsabs.harvard.edu/abs/1981ApJ...245.1035Y}
}

@ARTICLE{1986ApJ...305..740Z,
       author = {{Zhang}, E.-H. and {Robinson}, E.~L. and {Nather}, R.~E.},
        title = "{The Eclipses of Cataclysmic Variables. I. HT Cassiopeiae}",
      journal = {\apj},
         year = 1986,
        month = jun,
       volume = {305},
        pages = {740},
          doi = {10.1086/164288},
       adsurl = {https://ui.adsabs.harvard.edu/abs/1986ApJ...305..740Z}
}

@ARTICLE{1987AN....308...75W,
       author = {{Wenzel}, W.},
        title = "{The cycle length of the U Geminorum star HT Cassiopeiae}",
      journal = {Astronomische Nachrichten},
         year = 1987,
        month = jan,
       volume = {308},
        pages = {75},
          doi = {10.1002/asna.2113080115},
       adsurl = {https://ui.adsabs.harvard.edu/abs/1987AN....308...75W}
}

@ARTICLE{1996AJ....112.2248R,
       author = {{Robertson}, Jeff W. and {Honeycutt}, R. Kent},
        title = "{High-State/Low-State Photometric Behavior in the Quiescent Level of the Cataclysmic Variable HT Cassiopeiae}",
      journal = {\aj},
         year = 1996,
        month = nov,
       volume = {112},
        pages = {2248},
          doi = {10.1086/118177},
       adsurl = {https://ui.adsabs.harvard.edu/abs/1996AJ....112.2248R}
}

@ARTICLE{1999MNRAS.310..398I,
       author = {{Ioannou}, Zach and {Naylor}, T. and {Welsh}, W.~F. and {Catal{\'a}n}, M.~S. and {Worraker}, W.~J. and {James}, N.~D.},
        title = "{The `outside-in' outburst of HT Cassiopeiae}",
      journal = {\mnras},
         year = 1999,
        month = dec,
       volume = {310},
       number = {2},
        pages = {398-406},
          doi = {10.1046/j.1365-8711.1999.03001.x},
archivePrefix = {arXiv},
       eprint = {astro-ph/9907144},
 primaryClass = {astro-ph},
       adsurl = {https://ui.adsabs.harvard.edu/abs/1999MNRAS.310..398I}
}

@ARTICLE{2005MNRAS.364.1158F,
       author = {{Feline}, W.~J. and {Dhillon}, V.~S. and {Marsh}, T.~R. and {Watson}, C.~A. and {Littlefair}, S.~P.},
        title = "{ULTRACAM photometry of the eclipsing cataclysmic variables GY Cnc, IR Com and HT Cas}",
      journal = {\mnras},
         year = 2005,
        month = dec,
       volume = {364},
       number = {4},
        pages = {1158-1167},
          doi = {10.1111/j.1365-2966.2005.09668.x},
archivePrefix = {arXiv},
       eprint = {astro-ph/0510438},
 primaryClass = {astro-ph},
       adsurl = {https://ui.adsabs.harvard.edu/abs/2005MNRAS.364.1158F}
}

@ARTICLE{1997ApJ...475..812M,
       author = {{Mukai}, Koji and {Wood}, Janet H. and {Naylor}, Tim and {Schlegel}, Eric M. and {Swank}, Jean H.},
        title = "{The X-Ray Eclipse of the Dwarf Nova HT Cassiopeiae: Results from ASCA and ROSAT HRI Observations}",
      journal = {\apj},
         year = 1997,
        month = feb,
       volume = {475},
       number = {2},
        pages = {812-822},
          doi = {10.1086/303571},
       adsurl = {https://ui.adsabs.harvard.edu/abs/1997ApJ...475..812M}
}

@ARTICLE{2000A&A...359..998B,
       author = {{Bruch}, A.},
        title = "{Studies of the flickering in cataclysmic variables. VI. The location of the flickering light source in HT Cassiopeiae, V2051 Ophiuchi, IP Pegasi and UX Ursae Majoris}",
      journal = {\aap},
         year = 2000,
        month = jul,
       volume = {359},
        pages = {998-1010},
       adsurl = {https://ui.adsabs.harvard.edu/abs/2000A&A...359..998B}
}

@ARTICLE{2008A&A...480..481B,
       author = {{Borges}, B.~W. and {Baptista}, R. and {Papadimitriou}, C. and {Giannakis}, O.},
        title = "{Cyclical period changes in HT Cassiopeiae: a difference between systems above and below the period gap}",
      journal = {\aap},
         year = 2008,
        month = mar,
       volume = {480},
       number = {2},
        pages = {481-487},
          doi = {10.1051/0004-6361:20078596},
archivePrefix = {arXiv},
       eprint = {0711.3660},
 primaryClass = {astro-ph},
       adsurl = {https://ui.adsabs.harvard.edu/abs/2008A&A...480..481B}
}

@ARTICLE{2024ApJ...966..155S,
       author = {{Schaefer}, Bradley E.},
        title = "{Evolutionary Period Changes for 52 Cataclysmic Variables, and the Failure for the Most-fundamental Prediction of the Magnetic Braking Model}",
      journal = {\apj},
         year = 2024,
        month = may,
       volume = {966},
       number = {2},
          eid = {155},
        pages = {155},
          doi = {10.3847/1538-4357/ad31a9},
archivePrefix = {arXiv},
       eprint = {2404.12525},
 primaryClass = {astro-ph.SR},
       adsurl = {https://ui.adsabs.harvard.edu/abs/2024ApJ...966..155S}
}

@ARTICLE{2024ApJ...972...33S,
       author = {{Souza}, Leandro and {Baptista}, Raymundo},
        title = "{Cyclical Period Changes in Cataclysmic Variables: A Statistical Study}",
      journal = {\apj},
         year = 2024,
        month = sep,
       volume = {972},
       number = {1},
          eid = {33},
        pages = {33},
          doi = {10.3847/1538-4357/ad6b0e},
archivePrefix = {arXiv},
       eprint = {2408.07850},
 primaryClass = {astro-ph.SR},
       adsurl = {https://ui.adsabs.harvard.edu/abs/2024ApJ...972...33S}
}

@ARTICLE{2021MNRAS.507..809E,
       author = {{Er}, Huseyin and {{\"O}zd{\"o}nmez}, Aykut and {Nasiroglu}, Ilham},
        title = "{New observations of the eclipsing binary system NY Vir and its candidate circumbinary planets}",
      journal = {\mnras},
         year = 2021,
        month = oct,
       volume = {507},
       number = {1},
        pages = {809-817},
          doi = {10.1093/mnras/stab2054},
archivePrefix = {arXiv},
       eprint = {2107.07003},
 primaryClass = {astro-ph.SR},
       adsurl = {https://ui.adsabs.harvard.edu/abs/2021MNRAS.507..809E}
}

@ARTICLE{2023MNRAS.526.4725O,
       author = {{{\"O}zd{\"o}nmez}, Aykut and {Er}, Huseyin and {Nasiroglu}, Ilham},
        title = "{Investigation on the orbital period variations of NN Ser: implications for the hypothetical planets, the Applegate mechanism, and the orbital stability}",
      journal = {\mnras},
         year = 2023,
        month = dec,
       volume = {526},
       number = {3},
        pages = {4725-4734},
          doi = {10.1093/mnras/stad3086},
archivePrefix = {arXiv},
       eprint = {2310.05465},
 primaryClass = {astro-ph.EP},
       adsurl = {https://ui.adsabs.harvard.edu/abs/2023MNRAS.526.4725O}
}

@ARTICLE{2024PASA...41...47E,
       author = {{Er}, Huseyin and {{\"O}zd{\"o}nmez}, Aykut and {Nasiroglu}, Ilham and {Kenger}, Muhammet Emir},
        title = "{Orbital period variation analysis of the HS 0705+6700 post-common envelope binary}",
      journal = {\pasa},
         year = 2024,
        month = sep,
       volume = {41},
          eid = {e047},
        pages = {e047},
          doi = {10.1017/pasa.2024.50},
archivePrefix = {arXiv},
       eprint = {2405.13616},
 primaryClass = {astro-ph.SR},
       adsurl = {https://ui.adsabs.harvard.edu/abs/2024PASA...41...47E}
}

@ARTICLE{2025AdSpR..76.1204E,
       author = {{Er}, Huseyin and {Ozdonmez}, Aykut and {Kenger}, M. Emir and {G{\"u}rbulak}, B. Batuhan and {Nasiroglu}, Ilham},
        title = "{Investigating orbital periodicity in HS 2231 + 2441 with extended observations}",
      journal = {Advances in Space Research},
         year = 2025,
        month = jul,
       volume = {76},
       number = {2},
        pages = {1204-1212},
          doi = {10.1016/j.asr.2025.05.009},
       adsurl = {https://ui.adsabs.harvard.edu/abs/2025AdSpR..76.1204E}
}

@ARTICLE{2025ApJ...994...67E,
       author = {{Er}, Huseyin and {Ozdonmez}, Aykut and {Kenger}, M. Emir and {Gurbulak}, B. Batuhan and {Nasiroglu}, Ilham and {Tekkesinoglu}, Murat and {Ege}, Ergun and {Karaman}, Nazli},
        title = "{Investigating the System Configuration of Kepler-451 through Orbital Period Variations: Dynamical and Magnetic Interpretations}",
      journal = {\apj},
         year = 2025,
        month = nov,
       volume = {994},
       number = {1},
          eid = {67},
        pages = {67},
          doi = {10.3847/1538-4357/ae1029},
       adsurl = {https://ui.adsabs.harvard.edu/abs/2025ApJ...994...67E}
}

@ARTICLE{1985MNRAS.214..475W,
       author = {{Wood}, J.~H. and {Irwin}, M.~J. and {Pringle}, J.~E.},
        title = "{Adigital technique for the separation of the eclipses of a white dwarf and an accretion disc.}",
      journal = {\mnras},
         year = 1985,
        month = jun,
       volume = {214},
        pages = {475-479},
          doi = {10.1093/mnras/214.4.475},
       adsurl = {https://ui.adsabs.harvard.edu/abs/1985MNRAS.214..475W}
}

@ARTICLE{2010PASP..122..935E,
       author = {{Eastman}, Jason and {Siverd}, Robert and {Gaudi}, B. Scott},
        title = "{Achieving Better Than 1 Minute Accuracy in the Heliocentric and Barycentric Julian Dates}",
      journal = {\pasp},
         year = 2010,
        month = aug,
       volume = {122},
       number = {894},
        pages = {935},
          doi = {10.1086/655938},
archivePrefix = {arXiv},
       eprint = {1005.4415},
 primaryClass = {astro-ph.IM},
       adsurl = {https://ui.adsabs.harvard.edu/abs/2010PASP..122..935E}
}

@ARTICLE{1952ApJ...116..211I,
       author = {{Irwin}, John B.},
        title = "{The Determination of a Light-Time Orbit.}",
      journal = {\apj},
         year = 1952,
        month = jul,
       volume = {116},
        pages = {211},
          doi = {10.1086/145604},
       adsurl = {https://ui.adsabs.harvard.edu/abs/1952ApJ...116..211I}
}

@ARTICLE{2017AJ....153..137N,
       author = {{Nasiroglu}, Ilham and {Go{\'z}dziewski}, Krzysztof and {S{\l}owikowska}, Aga and {Krzeszowski}, Krzysztof and {{\.Z}ejmo}, Micha{\l} and {Zola}, Staszek and {Er}, Huseyin and {Og{\l}oza}, Waldemar and {Dr{\'o}{\.z}d{\.z}}, Marek and {Koziel-Wierzbowska}, Dorota and {Debski}, Bartlomiej and {Karaman}, Nazli},
        title = "{Is There a Circumbinary Planet around NSVS 14256825?}",
      journal = {\aj},
         year = 2017,
        month = mar,
       volume = {153},
       number = {3},
          eid = {137},
        pages = {137},
          doi = {10.3847/1538-3881/aa5d10},
archivePrefix = {arXiv},
       eprint = {1701.05211},
 primaryClass = {astro-ph.EP},
       adsurl = {https://ui.adsabs.harvard.edu/abs/2017AJ....153..137N}
}

@ARTICLE{2013PASP..125..306F,
       author = {{Foreman-Mackey}, Daniel and {Hogg}, David W. and {Lang}, Dustin and {Goodman}, Jonathan},
        title = "{emcee: The MCMC Hammer}",
      journal = {\pasp},
         year = 2013,
        month = mar,
       volume = {125},
       number = {925},
        pages = {306},
          doi = {10.1086/670067},
archivePrefix = {arXiv},
       eprint = {1202.3665},
 primaryClass = {astro-ph.IM},
       adsurl = {https://ui.adsabs.harvard.edu/abs/2013PASP..125..306F}
}

@ARTICLE{1984AcA....34..161S,
       author = {{Smak}, J.},
        title = "{Accretion in cataclysmic binaries. IV. Accretion disks in dwarf novae.}",
      journal = {\actaa},
         year = 1984,
        month = jan,
       volume = {34},
        pages = {161-189},
       adsurl = {https://ui.adsabs.harvard.edu/abs/1984AcA....34..161S}
}

@ARTICLE{1985MNRAS.213..129H,
       author = {{Horne}, K.},
        title = "{Images of accretion discs -I. The eclipse mapping method.}",
      journal = {\mnras},
         year = 1985,
        month = mar,
       volume = {213},
        pages = {129-141},
          doi = {10.1093/mnras/213.2.129},
       adsurl = {https://ui.adsabs.harvard.edu/abs/1985MNRAS.213..129H}
}

@ARTICLE{2002PASJ...54...79K,
       author = {{Kato}, Taichi and {Baba}, Hajime and {Nogami}, Daisaku},
        title = "{IR Com: Deeply Eclipsing Dwarf Nova Below the Period Gap -- A Twin of HT Cas?}",
      journal = {\pasj},
         year = 2002,
        month = feb,
       volume = {54},
        pages = {79-85},
          doi = {10.1093/pasj/54.1.79},
archivePrefix = {arXiv},
       eprint = {astro-ph/0110206},
 primaryClass = {astro-ph},
       adsurl = {https://ui.adsabs.harvard.edu/abs/2002PASJ...54...79K}
}

@article{1976Ap&SS..39..447L,
    author  = "Lomb, N.~R.",
    title   = "Least-Squares Frequency Analysis of Unequally Spaced Data",
    journal = "Astrophysics and Space Science",
    volume  = "39",
    number  = "2",
    year    = "1976",
    pages   = "447--462"
}

@article{1982ApJ...263..835S,
    author  = "Scargle, J.~D.",
    title   = "Studies in astronomical time series analysis. II. Statistical aspects of spectral analysis of unevenly spaced data",
    journal = "Astrophysical Journal",
    volume  = "263",
    number  = "",
    year    = "1982",
    pages   = "835--853"
}

@ARTICLE{2009Ap&SS.319..119T,
       author = {{Tian}, Y.~P. and {Xiang}, F.~Y. and {Tao}, X.},
        title = "{Period investigation of two RS CVn-type binary stars: RU Cancri and AW Herculis}",
      journal = {\apss},
         year = 2009,
        month = feb,
       volume = {319},
       number = {2-4},
        pages = {119-124},
          doi = {10.1007/s10509-008-9975-4},
       adsurl = {https://ui.adsabs.harvard.edu/abs/2009Ap&SS.319..119T}
}

@ARTICLE{2020MNRAS.491.1820L,
       author = {{Lanza}, A.~F.},
        title = "{Internal magnetic fields, spin-orbit coupling, and orbital period modulation in close binary systems}",
      journal = {\mnras},
         year = 2020,
        month = jan,
       volume = {491},
       number = {2},
        pages = {1820-1831},
          doi = {10.1093/mnras/stz3135},
archivePrefix = {arXiv},
       eprint = {1911.01757},
 primaryClass = {astro-ph.SR},
       adsurl = {https://ui.adsabs.harvard.edu/abs/2020MNRAS.491.1820L}
}

@ARTICLE{1990ApJ...357..621M,
       author = {{Marsh}, T.~R.},
        title = "{Detection of the Secondary Star in HT Cassiopeiae}",
      journal = {\apj},
         year = 1990,
        month = jul,
       volume = {357},
        pages = {621},
          doi = {10.1086/168950},
       adsurl = {https://ui.adsabs.harvard.edu/abs/1990ApJ...357..621M}
}

@ARTICLE{2012A&A...537A.128R,
       author = {{Rein}, H. and {Liu}, S. -F.},
        title = "{REBOUND: an open-source multi-purpose N-body code for collisional dynamics}",
      journal = {\aap},
         year = 2012,
        month = jan,
       volume = {537},
          eid = {A128},
        pages = {A128},
          doi = {10.1051/0004-6361/201118085},
archivePrefix = {arXiv},
       eprint = {1110.4876},
 primaryClass = {astro-ph.EP},
       adsurl = {https://ui.adsabs.harvard.edu/abs/2012A&A...537A.128R}
}

@ARTICLE{2015MNRAS.452..376R,
       author = {{Rein}, Hanno and {Tamayo}, Daniel},
        title = "{WHFAST: a fast and unbiased implementation of a symplectic Wisdom-Holman integrator for long-term gravitational simulations}",
      journal = {\mnras},
         year = 2015,
        month = sep,
       volume = {452},
       number = {1},
        pages = {376-388},
          doi = {10.1093/mnras/stv1257},
 primaryClass = {astro-ph.EP},
       adsurl = {https://ui.adsabs.harvard.edu/abs/2015MNRAS.452..376R}
}

@ARTICLE{2000A&AS..147..205C,
       author = {{Cincotta}, P.~M. and {Sim{\'o}}, C.},
        title = "{Simple tools to study global dynamics in non-axisymmetric galactic potentials - I}",
      journal = {\aaps},
         year = 2000,
        month = dec,
       volume = {147},
        pages = {205-228},
          doi = {10.1051/aas:2000108},
       adsurl = {https://ui.adsabs.harvard.edu/abs/2000A&AS..147..205C}
}

@ARTICLE{2021MNRAS.506.2122B,
       author = {{Brown-Sevilla}, S.~B. and {Nascimbeni}, V. and {Borsato}, L. and {Tartaglia}, L. and {Nardiello}, D. and {Granata}, V. and {Libralato}, M. and {Damasso}, M. and {Piotto}, G. and {Pollacco}, D. and {West}, R.~G. and {Colombo}, L.~S. and {Cunial}, A. and {Piazza}, G. and {Scaggiante}, F.},
        title = "{A new photometric and dynamical study of the eclipsing binary star HW Virginis}",
      journal = {\mnras},
         year = 2021,
        month = sep,
       volume = {506},
       number = {2},
        pages = {2122-2135},
          doi = {10.1093/mnras/stab1843},
 primaryClass = {astro-ph.SR},
       adsurl = {https://ui.adsabs.harvard.edu/abs/2021MNRAS.506.2122B}
}

@ARTICLE{1999AJ....117..621H,
       author = {{Holman}, Matthew J. and {Wiegert}, Paul A.},
        title = "{Long-Term Stability of Planets in Binary Systems}",
      journal = {\aj},
         year = 1999,
        month = jan,
       volume = {117},
       number = {1},
        pages = {621-628},
          doi = {10.1086/300695},
archivePrefix = {arXiv},
       eprint = {astro-ph/9809315},
 primaryClass = {astro-ph},
       adsurl = {https://ui.adsabs.harvard.edu/abs/1999AJ....117..621H}
}

@ARTICLE{2025ARep...69..758K,
       author = {{Kenger}, M.~E. and {Er}, H. and {{\"O}zd{\"o}nmez}, A.},
        title = "{Dynamical Orbital Stability Analyses of Eclipsing Binaries with Additional Companions}",
      journal = {Astronomy Reports},
         year = 2025,
        month = aug,
       volume = {69},
       number = {8},
        pages = {758-765},
          doi = {10.1134/S1063772924600444},
       adsurl = {https://ui.adsabs.harvard.edu/abs/2025ARep...69..758K}
}

@ARTICLE{2014ApJ...788...48S,
       author = {{Shappee}, B.~J. and {Prieto}, J.~L. and {Grupe}, D. and {Kochanek}, C.~S. and {Stanek}, K.~Z. and {De Rosa}, G. and {Mathur}, S. and {Zu}, Y. and {Peterson}, B.~M. and {Pogge}, R.~W. and {Komossa}, S. and {Im}, M. and {Jencson}, J. and {Holoien}, T.~W.-S. and {Basu}, U. and {Beacom}, J.~F. and {Szczygie{\l}}, D.~M. and {Brimacombe}, J. and {Adams}, S. and {Campillay}, A. and {Choi}, C. and {Contreras}, C. and {Dietrich}, M. and {Dubberley}, M. and {Elphick}, M. and {Foale}, S. and {Giustini}, M. and {Gonzalez}, C. and {Hawkins}, E. and {Howell}, D.~A. and {Hsiao}, E.~Y. and {Koss}, M. and {Leighly}, K.~M. and {Morrell}, N. and {Mudd}, D. and {Mullins}, D. and {Nugent}, J.~M. and {Parrent}, J. and {Phillips}, M.~M. and {Pojmanski}, G. and {Rosing}, W. and {Ross}, R. and {Sand}, D. and {Terndrup}, D.~M. and {Valenti}, S. and {Walker}, Z. and {Yoon}, Y.},
        title = "{The Man behind the Curtain: X-Rays Drive the UV through NIR Variability in the 2013 Active Galactic Nucleus Outburst in NGC 2617}",
      journal = {\apj},
         year = 2014,
        month = jun,
       volume = {788},
       number = {1},
          eid = {48},
        pages = {48},
          doi = {10.1088/0004-637X/788/1/48},
archivePrefix = {arXiv},
       eprint = {1310.2241},
 primaryClass = {astro-ph.HE},
       adsurl = {https://ui.adsabs.harvard.edu/abs/2014ApJ...788...48S}
}

@ARTICLE{2020MNRAS.491...13J,
       author = {{Jayasinghe}, T. and {Stanek}, K.~Z. and {Kochanek}, C.~S. and {Shappee}, B.~J. and {Holoien}, T.~W.-S. and {Thompson}, T.~A. and {Prieto}, J.~L. and {Dong}, S. and {Pawlak}, M. and {Pejcha}, O. and {Shields}, J.~V. and {Pojmanski}, G. and {Otero}, S. and {Hurst}, N. and {Britt}, C.~A. and {Will}, D.},
        title = "{The ASAS-SN catalogue of variable stars - V. Variables in the Southern hemisphere}",
      journal = {\mnras},
         year = 2020,
        month = jan,
       volume = {491},
       number = {1},
        pages = {13-28},
          doi = {10.1093/mnras/stz2711},
archivePrefix = {arXiv},
       eprint = {1907.10609},
 primaryClass = {astro-ph.SR},
       adsurl = {https://ui.adsabs.harvard.edu/abs/2020MNRAS.491...13J}
}

@ARTICLE{2012MNRAS.427.2812H,
       author = {{Horner}, J. and {Hinse}, T.~C. and {Wittenmyer}, R.~A. and {Marshall}, J.~P. and {Tinney}, C.~G.},
        title = "{A dynamical analysis of the proposed circumbinary HW Virginis planetary system}",
      journal = {\mnras},
         year = 2012,
        month = dec,
       volume = {427},
       number = {4},
        pages = {2812-2823},
          doi = {10.1111/j.1365-2966.2012.22046.x},
archivePrefix = {arXiv},
       eprint = {1209.0608},
 primaryClass = {astro-ph.EP},
       adsurl = {https://ui.adsabs.harvard.edu/abs/2012MNRAS.427.2812H}
}

@ARTICLE{2010MNRAS.408..631N,
       author = {{Nordhaus}, J. and {Spiegel}, D.~S. and {Ibgui}, L. and {Goodman}, J. and {Burrows}, A.},
        title = "{Tides and tidal engulfment in post-main-sequence binaries: period gaps for planets and brown dwarfs around white dwarfs}",
      journal = {\mnras},
         year = 2010,
        month = oct,
       volume = {408},
       number = {1},
        pages = {631-641},
          doi = {10.1111/j.1365-2966.2010.17155.x},
archivePrefix = {arXiv},
       eprint = {1002.2216},
 primaryClass = {astro-ph.SR},
       adsurl = {https://ui.adsabs.harvard.edu/abs/2010MNRAS.408..631N}
}

@ARTICLE{2001A&A...378..569G,
       author = {{Go{\'z}dziewski}, K. and {Bois}, E. and {Maciejewski}, A.~J. and {Kiseleva-Eggleton}, L.},
        title = "{Global dynamics of planetary systems with the MEGNO criterion}",
      journal = {\aap},
         year = 2001,
        month = nov,
       volume = {378},
        pages = {569-586},
          doi = {10.1051/0004-6361:20011189},
       adsurl = {https://ui.adsabs.harvard.edu/abs/2001A&A...378..569G}
}

@ARTICLE{2006OEJV...23...13P,
       author = {{Paschke}, A. and {Brat}, L.},
        title = "{O-C Gateway, a Collection of Minima Timings}",
      journal = {Open European Journal on Variable Stars},
         year = 2006,
        month = feb,
       volume = {23},
        pages = {13},
       adsurl = {https://ui.adsabs.harvard.edu/abs/2006OEJV...23...13P}
}

@ARTICLE{2013A&A...549A..95Z,
       author = {{Zorotovic}, M. and {Schreiber}, M.~R.},
        title = "{Origin of apparent period variations in eclipsing post-common-envelope binaries}",
      journal = {\aap},
         year = 2013,
        month = jan,
       volume = {549},
          eid = {A95},
        pages = {A95},
          doi = {10.1051/0004-6361/201220321},
archivePrefix = {arXiv},
       eprint = {1211.5356},
 primaryClass = {astro-ph.SR},
       adsurl = {https://ui.adsabs.harvard.edu/abs/2013A&A...549A..95Z}
}

@ARTICLE{2015AN....336..458S,
       author = {{Schleicher}, D.~R.~G. and {Dreizler}, S. and {V{\"o}lschow}, M. and {Banerjee}, R. and {Hessman}, F.~V.},
        title = "{Planet formation in post-common-envelope binaries}",
      journal = {Astronomische Nachrichten},
         year = 2015,
        month = jun,
       volume = {336},
       number = {5},
        pages = {458},
          doi = {10.1002/asna.201412184},
archivePrefix = {arXiv},
       eprint = {1501.01656},
 primaryClass = {astro-ph.SR},
       adsurl = {https://ui.adsabs.harvard.edu/abs/2015AN....336..458S}
}

@ARTICLE{2016RSOS....350571V,
       author = {{Veras}, Dimitri},
        title = "{Post-main-sequence planetary system evolution}",
      journal = {Royal Society Open Science},
         year = 2016,
        month = feb,
       volume = {3},
          eid = {150571},
        pages = {150571},
          doi = {10.1098/rsos.150571},
archivePrefix = {arXiv},
       eprint = {1601.05419},
 primaryClass = {astro-ph.EP},
       adsurl = {https://ui.adsabs.harvard.edu/abs/2016RSOS....350571V}
}

@ARTICLE{2026AdSpR..77.1365E,
       author = {{Er}, Huseyin and {Ozdonmez}, Aykut and {Kenger}, M. Emir and {Tekkesinoglu}, Murat and {Ege}, Ergun},
        title = "{Study of orbital period variation in DK Cyg using extended eclipse timing data}",
      journal = {Advances in Space Research},
         year = 2026,
        month = jan,
       volume = {77},
       number = {1},
        pages = {1365-1371},
          doi = {10.1016/j.asr.2025.10.029},
       adsurl = {https://ui.adsabs.harvard.edu/abs/2026AdSpR..77.1365E}
}

@ARTICLE{2013MNRAS.435.2033H,
       author = {{Horner}, J. and {Wittenmyer}, R.~A. and {Hinse}, T.~C. and {Marshall}, J.~P. and {Mustill}, A.~J. and {Tinney}, C.~G.},
        title = "{A detailed dynamical investigation of the proposed QS Virginis planetary system}",
      journal = {\mnras},
         year = 2013,
        month = nov,
       volume = {435},
       number = {3},
        pages = {2033-2039},
          doi = {10.1093/mnras/stt1420},
archivePrefix = {arXiv},
       eprint = {1307.7893},
 primaryClass = {astro-ph.EP},
       adsurl = {https://ui.adsabs.harvard.edu/abs/2013MNRAS.435.2033H}
}

@ARTICLE{2010MNRAS.407.2362P,
       author = {{Parsons}, S.~G. and {Marsh}, T.~R. and {Copperwheat}, C.~M. and {Dhillon}, V.~S. and {Littlefair}, S.~P. and {Hickman}, R.~D.~G. and {Maxted}, P.~F.~L. and {G{\"a}nsicke}, B.~T. and {Unda-Sanzana}, E. and {Colque}, J.~P. and {Barraza}, N. and {S{\'a}nchez}, N. and {Monard}, L.~A.~G.},
        title = "{Orbital period variations in eclipsing post-common-envelope binaries}",
      journal = {\mnras},
         year = 2010,
        month = oct,
       volume = {407},
       number = {4},
        pages = {2362-2382},
          doi = {10.1111/j.1365-2966.2010.17063.x},
archivePrefix = {arXiv},
       eprint = {1005.3958},
 primaryClass = {astro-ph.SR},
       adsurl = {https://ui.adsabs.harvard.edu/abs/2010MNRAS.407.2362P}
}

@ARTICLE{2011MNRAS.416.2202P,
       author = {{Potter}, Stephen B. and {Romero-Colmenero}, Encarni and {Ramsay}, Gavin and {Crawford}, Steven and {Gulbis}, Amanda and {Barway}, Sudhanshu and {Zietsman}, Ewald and {Kotze}, Marissa and {Buckley}, David A.~H. and {O'Donoghue}, Darragh and {Siegmund}, O.~H.~W. and {McPhate}, J. and {Welsh}, B.~Y. and {Vallerga}, John},
        title = "{Possible detection of two giant extrasolar planets orbiting the eclipsing polar UZ Fornacis}",
      journal = {\mnras},
         year = 2011,
        month = sep,
       volume = {416},
       number = {3},
        pages = {2202-2211},
          doi = {10.1111/j.1365-2966.2011.19198.x},
archivePrefix = {arXiv},
       eprint = {1106.1404},
 primaryClass = {astro-ph.SR},
       adsurl = {https://ui.adsabs.harvard.edu/abs/2011MNRAS.416.2202P}
}

@ARTICLE{2012AJ....144...34H,
       author = {{Hinse}, Tobias Cornelius and {Go{\'z}dziewski}, Krzysztof and {Lee}, Jae Woo and {Haghighipour}, Nader and {Lee}, Chung-Uk},
        title = "{The Proposed Quadruple System SZ Herculis: Revised LITE Model and Orbital Stability Study}",
      journal = {\aj},
         year = 2012,
        month = aug,
       volume = {144},
       number = {2},
          eid = {34},
        pages = {34},
          doi = {10.1088/0004-6256/144/2/34},
archivePrefix = {arXiv},
       eprint = {1202.4817},
 primaryClass = {astro-ph.SR},
       adsurl = {https://ui.adsabs.harvard.edu/abs/2012AJ....144...34H}
}

@ARTICLE{2022ExA....53...45A,
       author = {{Azzam}, Yosry A. and {Elnagahy}, F.~I.~Y. and {Ali}, Gamal B. and {Essam}, A. and {Saad}, Somaya and {Ismail}, Hamed and {Zead}, I. and {Ahmed}, Nasser M. and {Yoshida}, Michitoshi and {Kawabata}, Koji S. and {Akitaya}, Hiroshi and {Shokry}, A. and {Hendy}, Y.~H.~M. and {Takey}, Ali and {Hamed}, G.~M. and {Mack}, Peter},
        title = "{Kottamia Faint Imaging Spectro-Polarimeter (KFISP): opto-mechanical design, software control and performance analysis}",
      journal = {Experimental Astronomy},
         year = 2022,
        month = feb,
       volume = {53},
       number = {1},
        pages = {45-70},
          doi = {10.1007/s10686-021-09802-z},
       adsurl = {https://ui.adsabs.harvard.edu/abs/2022ExA....53...45A}
}

@ARTICLE{2016MNRAS.460.3873B,
       author = {{Bours}, M.~C.~P. and {Marsh}, T.~R. and {Parsons}, S.~G. and {Dhillon}, V.~S. and {Ashley}, R.~P. and {Bento}, J.~P. and {Breedt}, E. and {Butterley}, T. and {Caceres}, C. and {Chote}, P. and {Copperwheat}, C.~M. and {Hardy}, L.~K. and {Hermes}, J.~J. and {Irawati}, P. and {Kerry}, P. and {Kilkenny}, D. and {Littlefair}, S.~P. and {McAllister}, M.~J. and {Rattanasoon}, S. and {Sahman}, D.~I. and {Vu{\v{c}}kovi{\'c}}, M. and {Wilson}, R.~W.},
        title = "{Long-term eclipse timing of white dwarf binaries: an observational hint of a magnetic mechanism at work}",
      journal = {\mnras},
         year = 2016,
        month = aug,
       volume = {460},
       number = {4},
        pages = {3873-3887},
          doi = {10.1093/mnras/stw1203},
archivePrefix = {arXiv},
       eprint = {1606.00780},
 primaryClass = {astro-ph.SR},
       adsurl = {https://ui.adsabs.harvard.edu/abs/2016MNRAS.460.3873B}
}

@ARTICLE{2026MNRAS.547ag358Y,
       author = {{Yates}, Amalie and {Parsons}, S.~G. and {Brown}, A.~J. and {Castro Segura}, N. and {Dhillon}, V.~S. and {Dyer}, M.~J. and {Garbutt}, J.~A. and {Green}, M.~J. and {Jarvis}, D. and {Kennedy}, M.~R. and {Kerry}, P. and {Kilkenny}, D. and {Littlefair}, S.~P. and {McCormac}, J. and {Munday}, J. and {Pelisoli}, I. and {Pike}, E. and {Sahman}, D.~I.},
        title = "{Long-term eclipse time variations in white dwarf binaries}",
      journal = {\mnras},
         year = 2026,
        month = apr,
       volume = {547},
       number = {2},
          eid = {stag358},
        pages = {stag358},
          doi = {10.1093/mnras/stag358},
archivePrefix = {arXiv},
       eprint = {2602.17800},
 primaryClass = {astro-ph.SR},
       adsurl = {https://ui.adsabs.harvard.edu/abs/2026MNRAS.547ag358Y}
}

@ARTICLE{2026arXiv260427609N,
       author = {{Navarrete}, Felipe H. and {Schleicher}, Dominik R.~G. and {K{\"a}pyl{\"a}}, Petri J. and {V{\"o}lschow}, Marcel},
        title = "{Eclipsing time variations in close binaries produced by azimuthal dynamo waves}",
      journal = {arXiv e-prints},
         year = 2026,
        month = apr,
          eid = {arXiv:2604.27609},
        pages = {arXiv:2604.27609},
          doi = {10.48550/arXiv.2604.27609},
archivePrefix = {arXiv},
       eprint = {2604.27609},
 primaryClass = {astro-ph.SR},
       adsurl = {https://ui.adsabs.harvard.edu/abs/2026arXiv260427609N}
}

@ARTICLE{2025MNRAS.544...24P,
       author = {{Pulley}, D. and {Sharp}, I.~D. and {Mallett}, J. and {von Harrach}, S.},
        title = "{Eclipse timing variations in seven post-common envelope binaries: an update and analysis of their recently proposed circumbinary models}",
      journal = {\mnras},
         year = 2025,
        month = nov,
       volume = {544},
       number = {1},
        pages = {24-35},
          doi = {10.1093/mnras/staf1636},
archivePrefix = {arXiv},
       eprint = {2507.06748},
 primaryClass = {astro-ph.SR},
       adsurl = {https://ui.adsabs.harvard.edu/abs/2025MNRAS.544...24P}
}

@ARTICLE{2023MNRAS.526.2241B,
       author = {{Baycroft}, Thomas A. and {Triaud}, Amaury H.~M.~J. and {Kervella}, Pierre},
        title = "{New evidence about HW Vir's circumbinary planets from Hipparcos-Gaia astrometry and a reanalysis of the eclipse timing variations using nested sampling}",
      journal = {\mnras},
         year = 2023,
        month = dec,
       volume = {526},
       number = {2},
        pages = {2241-2250},
          doi = {10.1093/mnras/stad2794},
archivePrefix = {arXiv},
       eprint = {2309.05716},
 primaryClass = {astro-ph.EP},
       adsurl = {https://ui.adsabs.harvard.edu/abs/2023MNRAS.526.2241B}
}

@ARTICLE{2026ApJ...998..155X,
       author = {{Xiao}, Guang-Yao and {Feng}, Fabo and {Wang}, Song and {Li}, Kai and {Rui}, Yicheng and {Duan}, Xiao-Wei},
        title = "{Detection of Dark Companions via The Combination of Eclipse Timing Variation, Hipparcos, and/or Gaia Astrometry: The Cases of V Puppis and CY Ari}",
      journal = {\apj},
         year = 2026,
        month = feb,
       volume = {998},
       number = {1},
          eid = {155},
        pages = {155},
          doi = {10.3847/1538-4357/ae314c},
archivePrefix = {arXiv},
       eprint = {2512.20087},
 primaryClass = {astro-ph.SR},
       adsurl = {https://ui.adsabs.harvard.edu/abs/2026ApJ...998..155X}
}

@ARTICLE{2024A&A...691A.115Z,
       author = {{Zervas}, K. and {Christopoulou}, P.-E.},
        title = "{NSVS 14256825: Period variation and orbital stability analysis of two possible substellar companions}",
      journal = {\aap},
         year = 2024,
        month = nov,
       volume = {691},
          eid = {A115},
        pages = {A115},
          doi = {10.1051/0004-6361/202450195},
archivePrefix = {arXiv},
       eprint = {2408.15358},
 primaryClass = {astro-ph.SR},
       adsurl = {https://ui.adsabs.harvard.edu/abs/2024A&A...691A.115Z}
}
\bibliographystyle{aasjournalv7}



\end{document}